\documentclass{article}
\usepackage{graphicx}
\usepackage{amsmath}
\def\simlt{\mathrel{\hbox{\rlap{\hbox{\lower4pt\hbox{$\sim$}}}
\hbox{$<$}}}}
\def\simgt{\mathrel{\hbox{\rlap{\hbox{\lower4pt\hbox{$\sim$}}}
\hbox{$>$}}}}

\def\lya{Ly-$\alpha$ }

\title{Type IIn supernovae at $z\sim2$ from archival data}

\author{Jeff Cooke$^{1}$, Mark Sullivan$^{2}$, Elizabeth
J. Barton$^{1}$, James S. Bullock$^{1}$,\\ Ray G. Carlberg$^{3}$, 
Avishay Gal-Yam$^{4}$, Erik Tollerud$^{1}$}
 
\begin{document}
\maketitle
\thispagestyle{empty}
$^{1}$ Center for Cosmology, Department of Physics and Astronomy, 
University of California, Irvine, CA 92697-4574, USA 
$^{2}$ Department of Physics, University of Oxford, Denys
Wilkinson Building, Keble Road, Oxford OX1 3RH, UK
$^{3}$ Department of Astronomy and Astrophysics, University of
Toronto, Toronto, ON M5S 3H4, Canada
$^{4}$ Benoziyo Center for Astrophysics, Weizmann Institute of Science,
76100 Rehovot, Israel

\begin{abstract}
Supernovae have been confirmed to redshift
$z\sim1.7$~\cite{poznanski07,riess01} for type Ia (thermonuclear
detonation of a white dwarf) and to
$z\sim0.7$~\cite{poznanski07,botticella08,dellavalle06,soderberg06}
for type II (collapse of the core of the star).  The subclass type
IIn~\cite{schlegel90} supernovae are
luminous~\cite{richardson02,smith07,smith08} core-collapse explosions
of massive stars~\cite{smith07,smith08,kotak06,galyam07} and, unlike
other types, are very bright in the
ultraviolet~\cite{fransson02,fransson05,cooke08,brown08}, which should
enable them to be found optically at redshifts $z\sim2$ and
higher~\cite{dahlen99,cooke08}.  In addition, the interaction of the
ejecta with circumstellar material creates strong, long-lived emission
lines that allow spectroscopic confirmation of many events of this
type at $z\sim2$ for $3-5$ years after explosion~\cite{cooke08}.  Here
we report three spectroscopically confirmed type IIn supernovae, at
redshifts $z=0.808, 2.013$ and $2.357$, detected in archival data
using a method~\cite{cooke08} designed to exploit these properties at
$z\sim2$.  Type IIn supernovae directly probe the formation of massive
stars at high redshift. The number found to date is consistent with
the expectations of a locally measured~\cite{salpeter55} stellar
initial mass function, but not with an evolving initial mass function
proposed to explain independent observations at low and high redshift.
\end{abstract}

The three $z\sim2$ type IIn supernovae are detected in the Deep
component of the Canada-France-Hawaii Telescope Legacy Survey
(CFHT-LS) which consists of four fields, each of area one
degree-squared, imaged over five years in five ($u^*g'r'i'z'$)
filters.  Our approach is to monitor $z\sim2$ galaxies over multiple
years and search for flux variations that meet criteria for
high-redshift type IIn supernovae.  Galaxies are identified using
efficient colour-selection
techniques~\cite{steidel03,steidel04,cooke05} tailored to the CFHT-LS
and spectroscopically tested using the Vi(r)mos-VLT Deep
Survey~\cite{lefevre03}.  The nightly images from a given year (a
$5-6$ month ``season'' of observations) are combined into
``seasonal-stacked'' images for each filter.  The stacked images are
sensitive to the slow ($\sim3-5$ months at $z\sim2$) photometric
evolution of high-redshift type IIn supernovae as a result of
cosmological time dilation, and to events $\sim$6 times fainter than
those detected by the Supernova Legacy Survey that searches the same
high-quality data using conventional techniques~\cite{neill06}.
Finally, we use the ``seasonal-stacked'' images for candidate
inspection and the nightly exposures to construct high-resolution
light curves (see Supplementary Information for more details).

The confirmed $z\sim2$ type IIn supernovae are the first three
supernovae candidates detected in the two fields over four seasons
(2003-2006) analysed to date.  Candidates must meet conservative
criteria that include (1) detection of $\ge3\sigma$ in the $g'-,r'-$,
and $i'-$band ``seasonal-stacked'' images, (2) flux variation in only
one season (the first and last seasons are disregarded for this
reason), (3) clean point-source detections in the subtracted images,
(4) a priority for events with host galaxy centroid offsets, and (5)
$g'r'i'$ light curves that exhibit flux rise times and decay rates
consistent with supernova profiles.  The criteria are designed to
prevent misidentification of active galactic nuclei (AGN; accreting
supermassive black holes at the centres of galaxies) and other
contaminants that can mimic supernova events to a certain extent.

\begin{figure}
\begin{center} 
\scalebox{0.45}[0.45]{\rotatebox{0}{\includegraphics{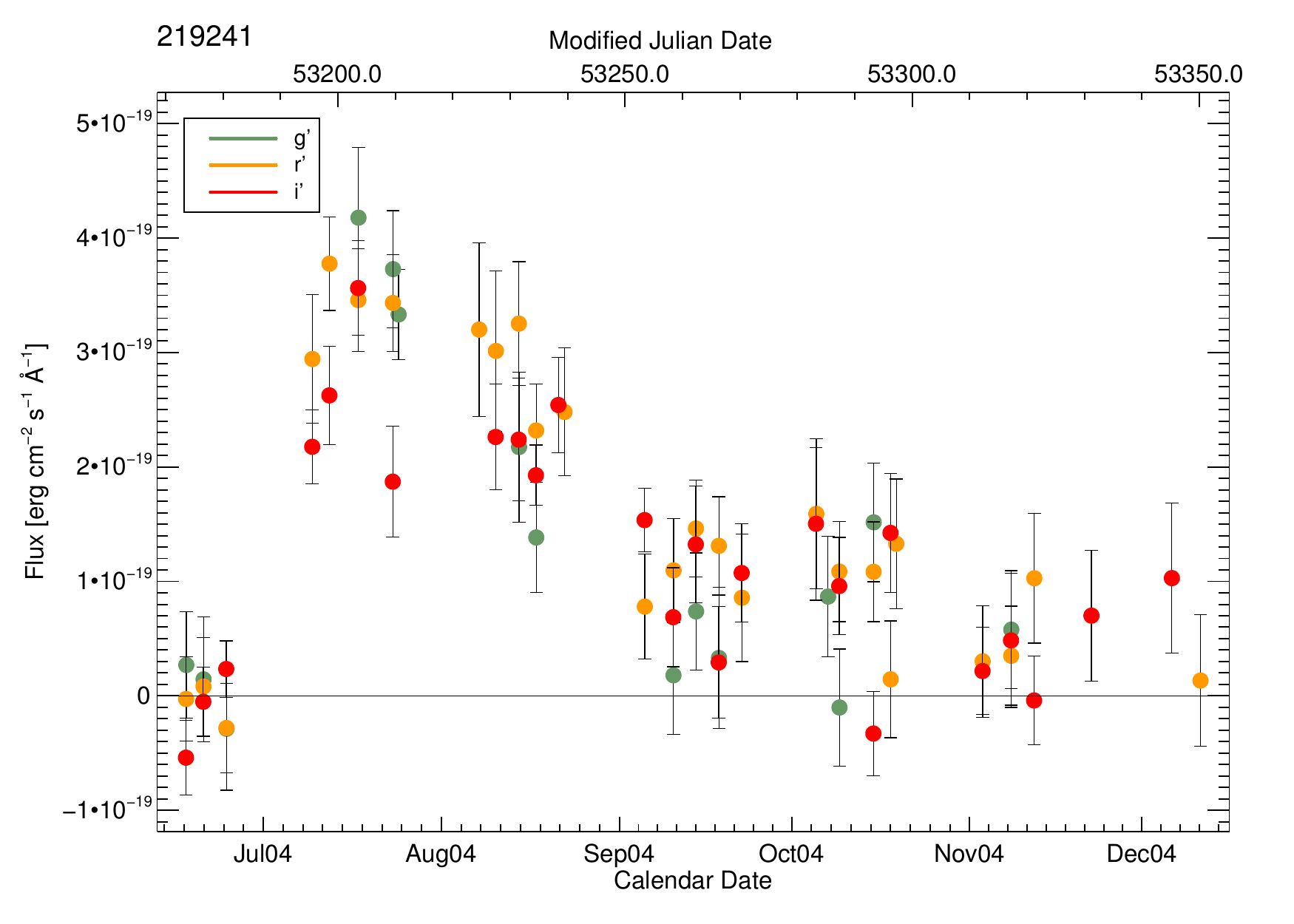}}}
\scalebox{0.45}[0.45]{\rotatebox{0}{\includegraphics{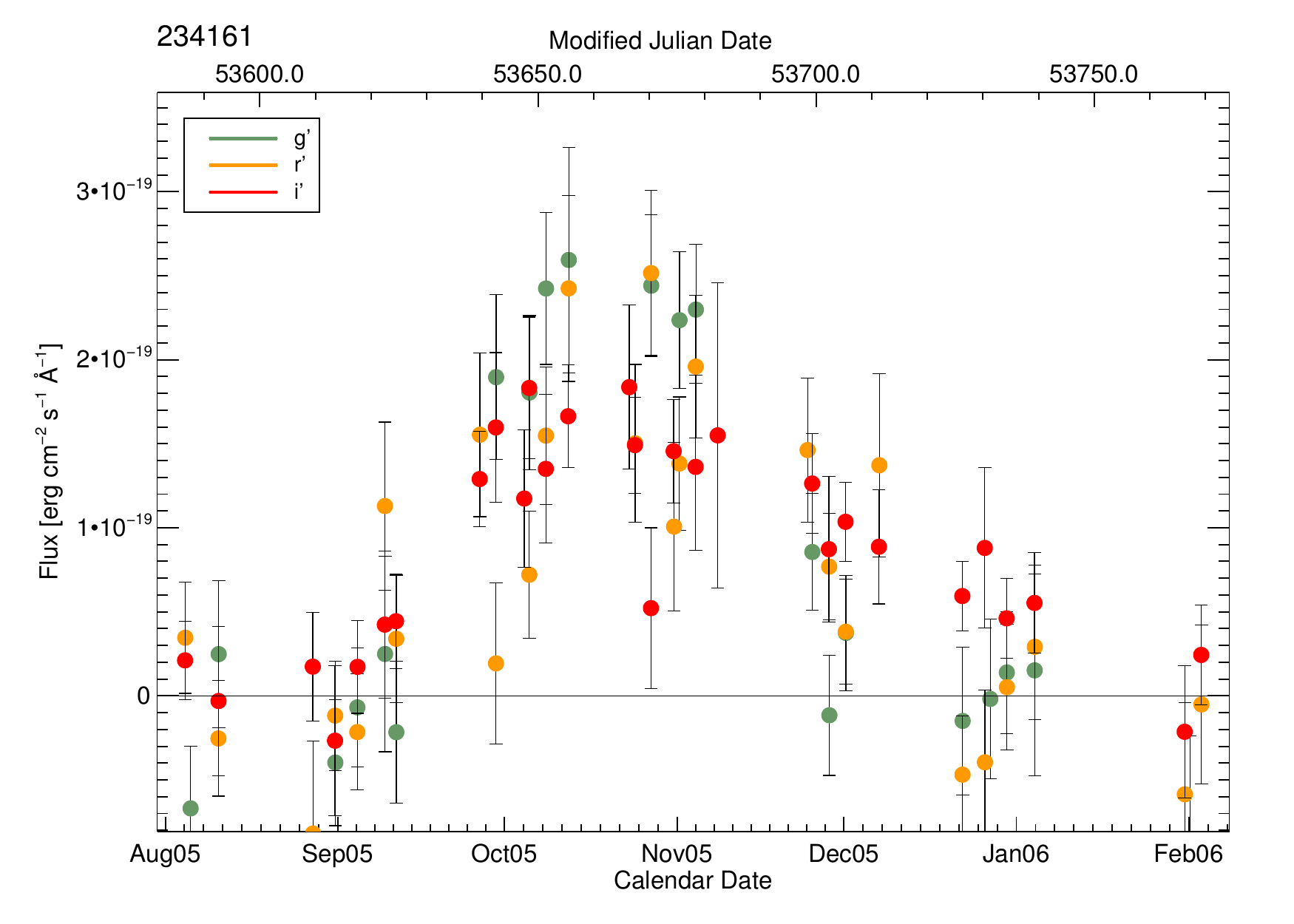}}}
\scalebox{0.45}[0.45]{\rotatebox{0}{\includegraphics{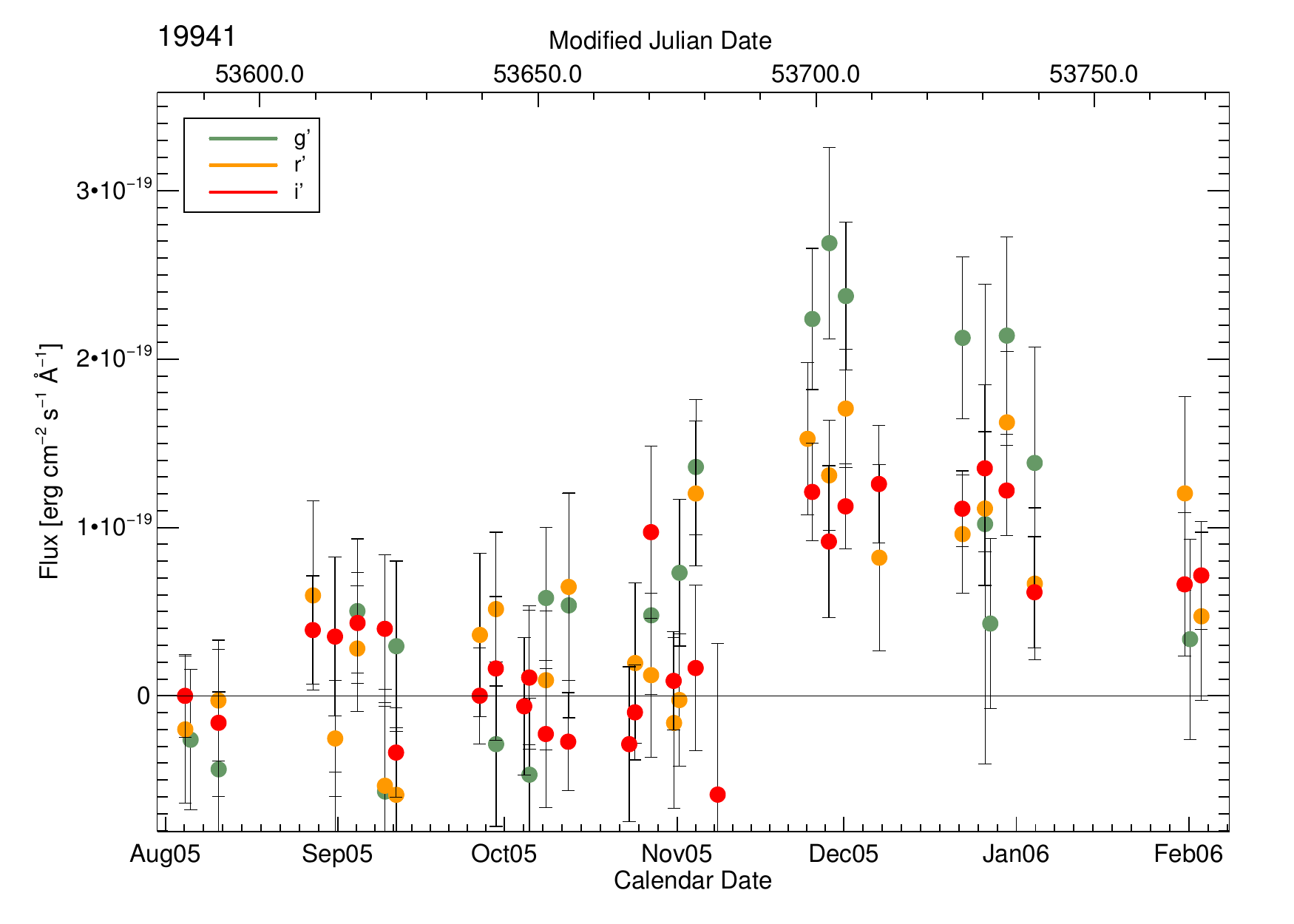}}}
\caption{\small Multi-colour light curves of the three high-redshift
 supernovae.  Top to bottom: SN 219241 ($z=0.808$), SN 234161
 ($z=2.013$), and SN 19941 ($z=2.357$).  The flux and $1\sigma$
 uncertainties for the $g'r'i'$ optical filters shown here probe the
 rest-frame ultraviolet flux for these events.  The seasonal
 integrated flux from each supernova is detected at $3-8.4\sigma$ over
 the host galaxy flux in each filter.  Because type Ia supernovae
 exhibit very little flux shortward of $\sim300$ nm as a result of
 efficient Fe\textsc{ii} scattering of ultraviolet
 photons~\cite{riess04}, this classification is ruled out.  We find
 that the ultraviolet luminosity and evolution of the supernovae in
 each filter are most consistent with type IIn events (see
 Supplementary Information).
\label{lc}}
\end{center}
\end{figure}

The $g'r'i'$ light curves for the three $z\sim2$ supernovae are
presented in Fig.~\ref{lc} and are photometrically consistent with
type IIn supernovae behaviour.  We confirmed the supernova redshifts
(Table~\ref{prop}) from deep spectroscopy obtained using the 10-m
telescopes at the W. M. Keck Observatory equipped with the Deep
Imaging Multi-Object Spectrograph~\cite{faber03} on 30 September 2008
and 01 October 2008 and with the Low Resolution Imaging
Spectrometer~\cite{oke95,mccarthy98} on 25 January 2009.  The ability
to spectroscopically confirm the events as type IIn supernovae via the
detection of strong, long-lived emission lines differs dramatically
from other SN types that require rapid follow-up to obtain spectral
classifications of quickly fading continua.

\begin{figure}
\begin{center} 
\scalebox{0.55}[0.55]{\rotatebox{90}{\includegraphics{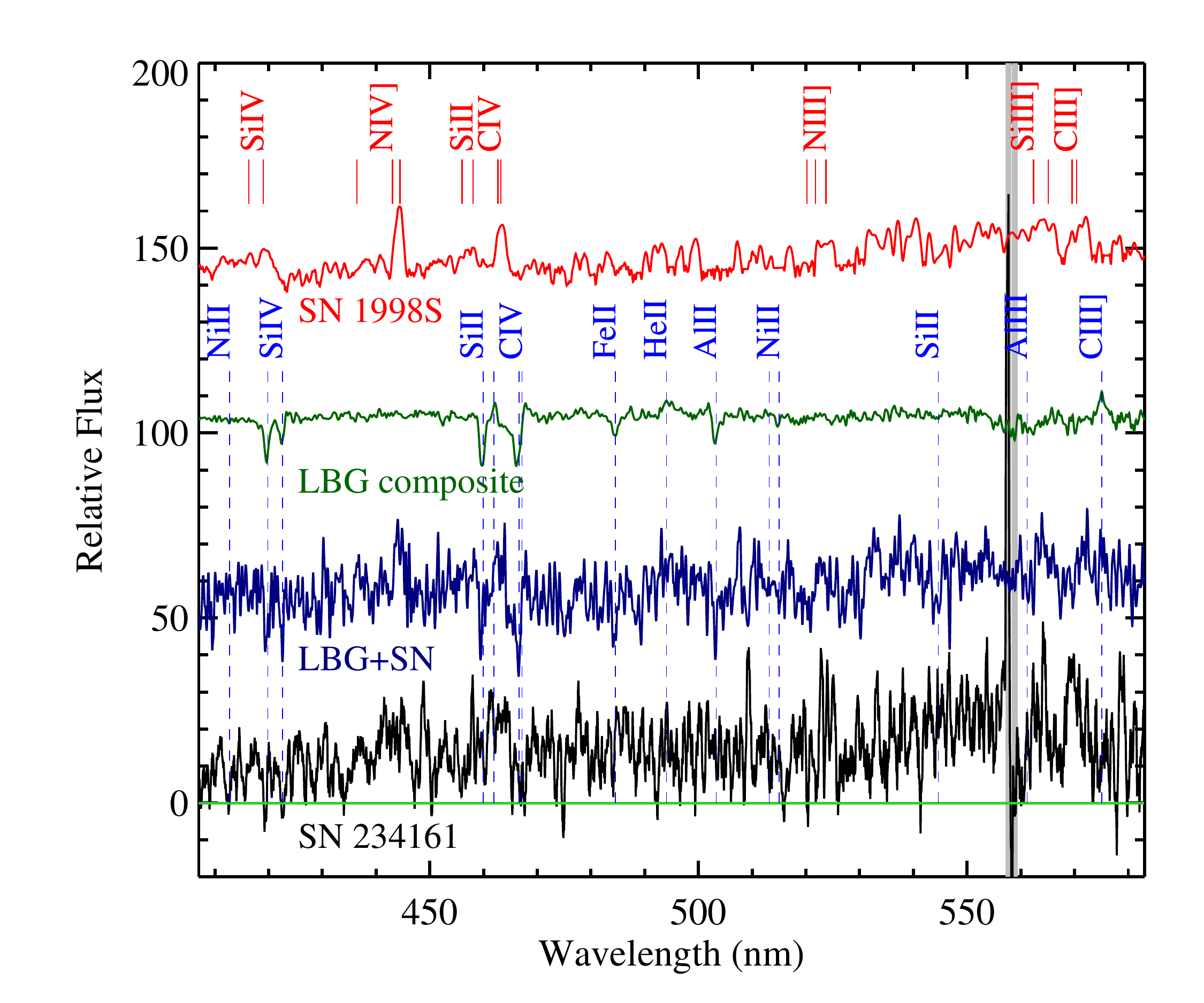}}}
\caption{\small Four spectra with vertical offsets to help illustrate
  the detection of line-emission from SN 234161.  {\bf ~(a)} The
  ultraviolet spectrum (red) of the low-redshift SN
  1998S~\cite{fransson05} (day 485) shifted and flux corrected to
  $z=2.013-2,500$ km s$^{-1}$ ($-2,500$ km s$^{-1}$ corresponds to
  $\Delta z=-0.026$; best fit match to SN 234161 emission).  The short
  solid vertical lines indicate the positions of expected type IIn
  supernovae emission-line features.  {\bf ~(b)} A composite $z\sim3$
  galaxy spectrum~\cite{shapley03} (green) representative of typical
  $z\sim2-3$ galaxies using our colour selection.  Dashed vertical
  lines indicate the expected positions of interstellar ultraviolet
  features.  {\bf ~(c)} Convolution of the composite galaxy and SN
  1998S spectra (blue) with Gaussian noise added consistent with that
  of the data.  {\bf ~(d)} The data for SN 234161 (day 365; black).
  The thick grey vertical line marks the position of a bright night
  sky emission line that is difficult to subtract cleanly from the
  faint spectrum.  The effect of the supernova emission on the
  convolved spectrum includes the Si\textsc{iv} 139.4, 140.3 nm
  profile, excess flux from N\textsc{iv]} 146.1, 148.3, 148.8 and
  N\textsc{iii]} 174.1, 174.7, 175.3 nm near 440 and 520 nm,
  respectively, excess flux and profile of C\textsc{iv} 154.8, 155.0
  nm near 460 nm, and the rise in the continuum beyond $\sim530$ nm.
  Comparison of the data to the convolved spectrum shows that SN
  234161 is consistent with an average $z\sim2-3$ galaxy experiencing
  a SN 1998S-like event with emission-line peaks blueshifted 2500 km
  sec$^{-1}$ (low-redshift type IIn supernovae also exhibit
  blue-shifted emission peaks).  In addition, SN 234161 exhibits
  strong C\textsc{iii]} 190.7, 190.9 nm supernova emission near 570
  nm.  We remark that SN 1998S and type IIn SN 1995N~\cite{fransson02}
  exhibited C\textsc{III]} emission of similar relative strength to SN
  234161 at earlier (day $238$) and later (day $943$) times,
  respectively.
\label{spec}}
\end{center}
\end{figure}

We detected late-time supernova emission to $>3\sigma$ significance in
the combined 6,000-s exposure of SN 234161 (365 days old, rest-frame;
Fig.~\ref{spec}).  The features are very similar to those seen at
low-redshift, and include ultraviolet shock ionization emission lines
(such as semi-forbidden transitions N\textsc{iv]} and N\textsc{iii]}
that are extremely rare in AGN), and emission-line strength ratio
values (for example, C\textsc{iv}/C\textsc{iii]}) that are not
theoretically predicted or observed in AGN.  In addition, the combined
5,400-s exposure of SN 19941 (345 days old, rest-frame) exhibits
strong blueshifted Si\textsc{iv} and weaker C\textsc{iv} emission
attributed to the supernova.  We did not detect significant
($>3\sigma$) supernova emission in the spectrum of SN 219241.  This is
probably a consequence of the supernova age (855 days old,
rest-frame), placing it at an epoch when the emission lines are
expected to fade to near or below the spectroscopic threshold of the
shorter combined 3,600-s exposure~\cite{cooke08}.  The detection and
eventual decay of type IIn supernova emission lines verify photometric
classification, and confirm the ability to study supernova energies
and chemistry on an individual basis to high redshift. (See
Supplementary Information for spectroscopic and line-emission
details.)

Progenitors of type IIn supernovae are believed to be massive
stars~\cite{kotak06,galyam07,smith07,smith08} that sample the
high-mass end of the stellar initial mass function (IMF) of galaxies.
Because this method colour-selects a well-controlled population of
galaxies over a well-defined volume, a small number of type IIn
supernovae can not only give the high-redshift rate of these
supernovae, but also provide the first direct probe of the high
redshift IMF.  For example, an evolving IMF model invoked recently to
reconcile indirect high- and low-redshift
observations~\cite{vandokkum08,dave08,chary08} predicts a greater
number (by a factor of $\sim3$) of $z\sim2$ type IIn supernovae than a
locally measured IMF~\cite{salpeter55}.  Computing the expectations
for the volume analysed to date, the confirmed type IIn supernovae
presented here are consistent with the relative number predicted using
a static local IMF.  A relaxation of our conservative supernova
criteria and a search for supernovae at greater radii from their host
centroids in the complete CFHT-LS dataset will test the validity of
this result.  Finally, the method presented here provides a means to
identify $\sim40,000$ type IIn supernovae at $z\sim2$ and to detect
events to $z\sim6$ over the next ten years with 8-m-class deep
synoptic optical campaigns, some of which are currently underway and
some of which are soon to begin.  This cannot be done with any other
supernova type or by conventional search practices.  As a result, the
exceptional properties of type IIn supernovae should enable a seamless
study of stellar and galactic processes, ranging from the local
Universe to a time shortly after the formation of the first stars.\\
\\

This work was made possible by the generous support provided by
the Gary McCue Postdoctoral Fellowship and the Center for Cosmology at
the University of California, Irvine.  We acknowledge support from
NSERC and the Royal Society.  The analysis pipeline used to reduce the
DEIMOS data was developed at UC Berkeley with support from NSF grant
AST-0071048.  The CFHT Legacy Survey relies on observations with
MegaCam, a joint project of CFHT and CEA/DAPNIA, at the
Canada-France-Hawaii Telescope (CFHT). We used data products from the
Canadian Astronomy Data Centre as part of the CFHT Legacy Survey.
Some of the data presented here were obtained at the W. M. Keck
Observatory.  Both observatories are located near the summit of Mauna
Kea, Hawai'i. The authors wish to recognize and acknowledge the very
significant cultural role and reverence that the summit of Mauna Kea
has always had within the indigenous Hawaiian community.  We are most
fortunate to have the opportunity to conduct observations from this
mountain.\\

The authors declare that they have no competing financial interests.\\

Correspondence should be addressed to Jeff Cooke~(email: cooke@uci.edu).

\clearpage

\begin{table}
\centering
{\scriptsize
\begin{tabular}{lcccccccc}
\hline SN & R.A. & Dec. & Host & Date of & SN & SN & Separation & Redshift\\
   & (J2000) & (J2000) & m$_r$ & Outburst & m$_r$ & M$_{UV}$ & kpc & $z$\\
\hline
 19941 & 02 24 11.147 & -04 57 58.41 & $25.16\pm0.07$ & 24Nov2005 &
  $25.6_{-0.3}^{+0.2}$ & $-19.5_{-0.3}^{+0.2}$ & $1.1\pm0.8$ & $2.357\pm0.002$\\
234161 & 02 24 33.271 & -04 26 31.04 & $24.93\pm0.07$ & 27Sep2005 &
  $25.2_{-0.2}^{+0.2}$ & $-19.6_{-0.2}^{+0.2}$ & $2.8\pm0.8$ & $2.013\pm0.002$\\
219241 & 22 14 47.351 & -17 46 11.62 & $24.12\pm0.04$ & 08Jul2004 &
  $24.7_{-0.1}^{+0.1}$ & $-18.2_{-0.1}^{+0.1}$ & $2.0\pm0.7$ & $0.808\pm0.001$\\\hline
\hline
\end{tabular}
\caption{{High-redshift supernova and host galaxy properties.  Both
the dates of outburst and the supernova (SN) magnitudes are determined
from the light curves.  Supernova apparent $r'$-band magnitudes
(m${r'}$) are at peak brightness and corresponding absolute
ultraviolet magnitudes ($M_{UV}$) are estimated at the $r'$-band
effective rest-frame wavelengths.  Separations are measured between
the centroids of the supernovae and their host galaxies and are
accurate to $\pm0.52$ pixels, which corresponds to a physical
separation of $0.7-0.8$ kpc (see Supplementary Information).}
\label{prop}} } 
\end{table}

\clearpage

\clearpage

\clearpage

\begin{center}
\Large{\bf Supplementary information}
\normalsize
\end{center}
\thispagestyle{empty}

Here we describe in more detail the image stacking process
(\S\ref{stacks}), the galaxy colour-selection technique
(\S\ref{colors}), and the supernova (SN) UV magnitudes and decay rates
derived from the images and light curves (\S\ref{images} \&
\ref{lightcurves}).  In addition, we present the spectroscopic
observations for the three $z\sim2$ SNe (\S\ref{spectra}) and discuss
evidence for SN line emission detection from SN 234161 and SN 19941
(\S\ref{emission}).

\section{Image stacks}
\label{stacks}

\hspace{1.0cm} ``Seasonal-stacked'' images in the two fields are made
from data from the individual nights observed in each of the $g'$,
$r'$, and $i'$ filters. A precise astrometric solution, accounting for
geometrical distortion, is assigned to every image frame, and the
images resampled to a common pixel coordinate system.  The seeing (the
FWHM of the point-spread function) and photometric quality are
determined by flux measurements of tertiary standard stars in the
CFHT-LS fields compiled by the SNLS team.  These tertiary standards
provide photometric zeropoints to the Vega system and are derived from
observations of secondary standard stars~\cite{landolt92}.
Two-dimensional sky variation is removed from each frame by fitting
the background spatially and subtracting the resultant fit.  Each
individual frame has a weight-map associated with it, containing the
uncertainty in each pixel from considerations of photon noise from the
sky background and object photons. Known bad pixels and saturated
pixels are assigned a weight of zero.  The stacks are then made from
the highest-quality data taken in each year as part of CFHT-LS: all
individual images entering the stacks have seeing $<0.75''$ FWHM.  The
resampled frames that pass this cut are combined using a weighted
average, with a $3\sigma$ clipping to remove artifacts such as
satellite trails and cosmic-rays. Typical exposure times for the final
stacked images range from 10,000 seconds (in $g'$) to 50,000 seconds
(in $i'$).

\section{Galaxy colour-selection}
\label{colors}

\hspace{1.0cm} We combine the ``seasonal-stacked'' images from all
four years into ``super-stacked'' images to colour-select
high-redshift galaxies.  In this manner we reach deeper than the
``seasonal-stacked'' images alone and are able determine galaxy
colours with photometric uncertainties $\le0.2$ mag for galaxies to
m$_{r'}\sim27$.  We detect sources in the field using {\it
SExtractor}~\cite{bertin96} software and use the segmentation map
generated from the ``super-stacked'' images to monitor the flux of
selected galaxies year-to-year in the ``seasonal-stacked'' images.
The higher sensitivity of the ``super-stacked'' images enable a more
complete pixel definition of the luminous extent of each galaxy for
optimal reference segmentation maps.  In addition, the combination of
four seasons effectively dilutes any potential colour contribution to
the host galaxy as a result of part or all of a SN event that may
affect the colour selection.  We note that $z\sim2$ SNe IIn are
detectable in the CFHT-LS data for $\sim3-5$ months, observed-frame.
Any significant contribution from a detectable SN would result in a
net flat or bluer addition the host galaxy colour in a given
``seasonal-stacked'' image and our colour criteria include the bluest
galaxies.  A somewhat redder SN contribution in the UV would come from
the fraction of an event caught after peak brightness which would in
turn be quite faint in integrated light in the particular
``seasonal-stacked'' image and thereby contribute very little to the
four-season ``super-stacked'' image.

We filter $z\ge1$ galaxies from the detected sources using a
colour-selection technique specifically designed for the CFHT MegaCam.
We determine the colours of 10 different galaxy templates evolved from
$z=0$ to high redshift by convolving the templates with (1) the
throughput of the CFHT MegaCam $u^*g'r'i'z'$ filters, (2) the MegaCam
CCD quantum efficiency, (3) the atmospheric extinction of Mauna Kea,
and (4) an adopted high-redshift prescription~\cite{madau95} for the
expected decrement of flux shortward of \lya and the Lyman limit as a
result of intervening systems and intrinsic absorption.  We use the
spectroscopic catalogue of the Vi(r)mos-VLT Deep Survey (VVDS)~$^{24}$
to test our CFHT-LS colour selection criteria and find that $83$\% of
the galaxies common to both surveys that have both VVDS
high-confidence redshift qualifiers and that meet our colour criteria
have redshifts $z>1$.  The colour cuts that isolate $z\sim2$ and
$z\sim3$ galaxies result in redshift distributions $\langle
z\rangle=1.68\pm0.34$ and $\langle z\rangle=3.17\pm0.41$, respectively
(there are too few galaxies for a meaningful distribution for our
$z>3$ colour cuts).  We monitor all sources that meet these color
criteria and consider a relaxed color cut that includes $z>1$ sources.

\section{Supernova photometric detection}
\label{images}

\hspace{1.0cm} SNe IIn have a magnitude distribution $\langle
M_B\rangle=-19.0~\pm 0.3,~\sigma=0.92$~$^7$ and UV luminosities
$\sim1.0$ magnitude fainter near peak
brightness~$^{13,}$\cite{immler06,immler07}.  SNe template analyses
show that $\sim93$\% of all SN detections at $z\ge2$ are expected to
be SNe IIn~$^{16,14}$ as a result of their strong intrinsic UV
luminosity.  We searched the CFHT-LS data for events are that meet the
criteria discussed in the main Letter that includes a $3\sigma$ flux
variation over the host galaxy flux in each of the $g'r'i'$ filters.
Of the initial four $z\sim2$ SN candidates, two SN events (SN 234161
and SN 19941) meet the $\langle z\rangle=1.68\pm0.34$ and $\langle
z\rangle=3.17\pm0.41$ colour-selection and the remaining two events
(SN 219241 and SN 352912) meet relaxed colour cuts that select
galaxies with redshifts above $z\sim1$.  Follow-up spectroscopy
(\S\ref{spectra}) of SN 234161 and SN 19941 confirmed host redshifts
of $z=2.0125$ and $z=2.3565$.  In addition, we have obtained
spectroscopy of SN 219241 and find the host redshift to be $z=0.8078$.
From the above results and the efficiency of our colour selection, we
expect the SN 352912 to have a high likelihood to be a $z\ge1$ event.

\begin{figure}
\begin{center}
\scalebox{0.21}[0.21]{\rotatebox{0}{\includegraphics{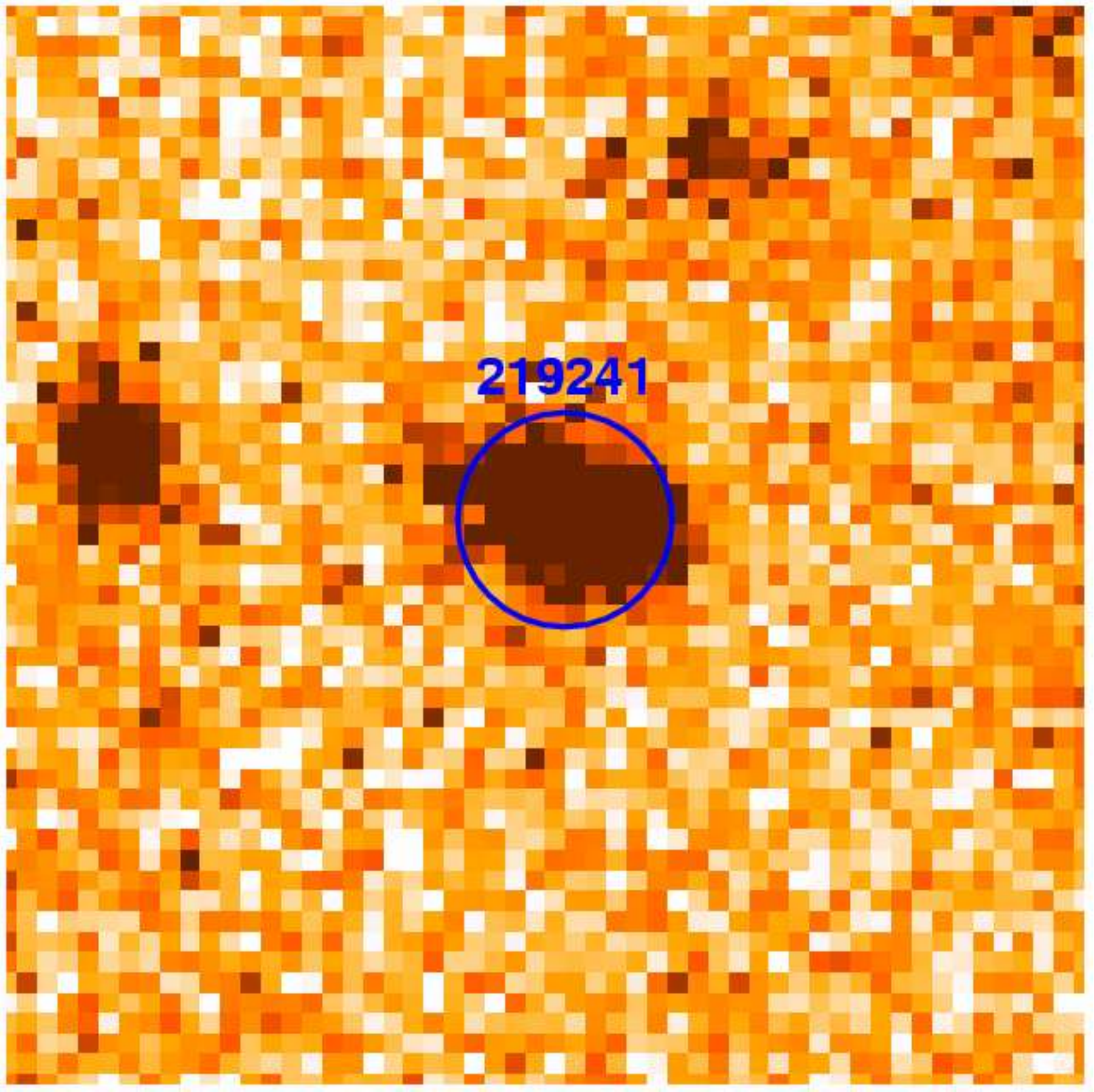}}}
\scalebox{0.21}[0.21]{\rotatebox{0}{\includegraphics{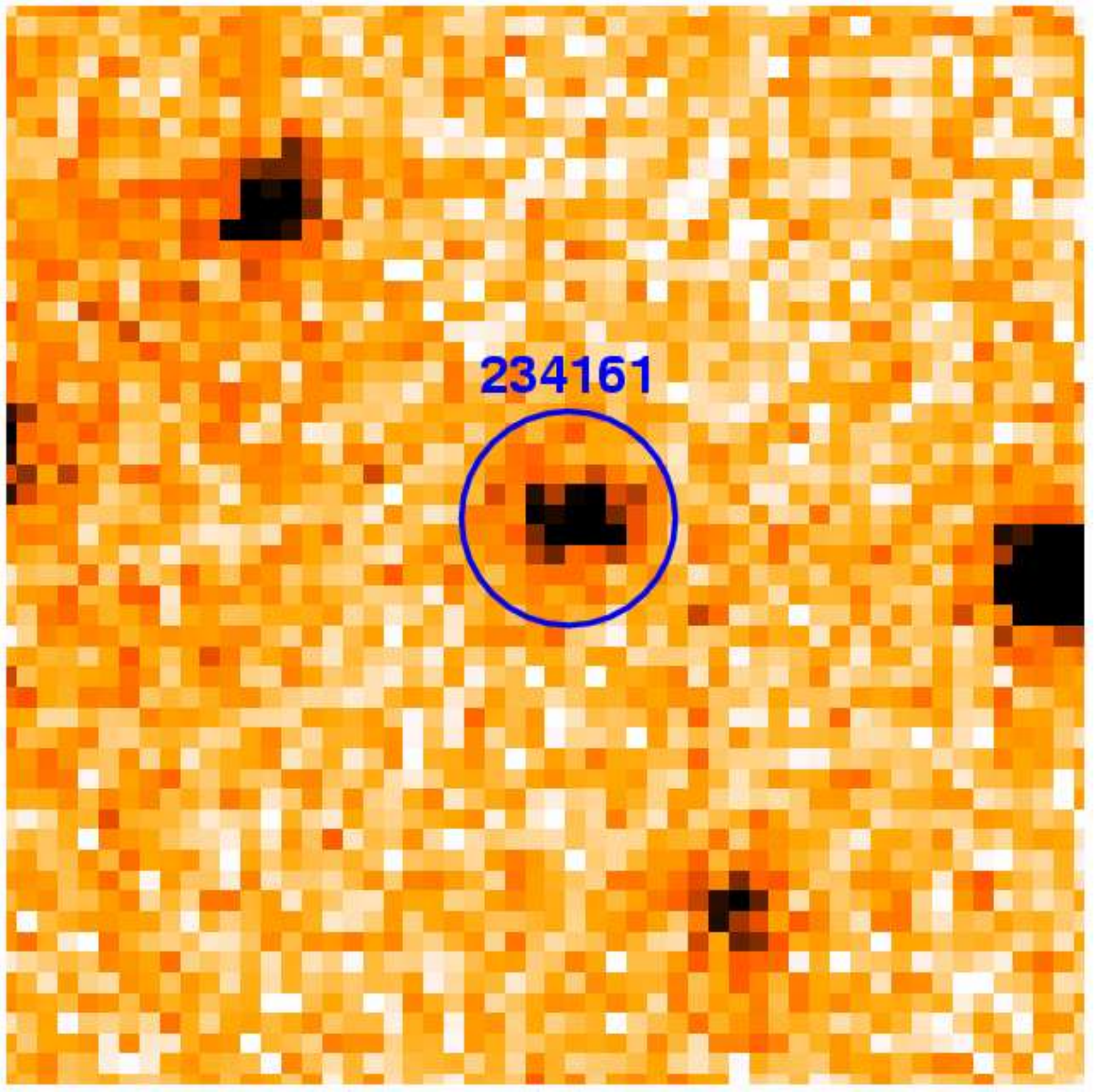}}}
\scalebox{0.21}[0.21]{\rotatebox{0}{\includegraphics{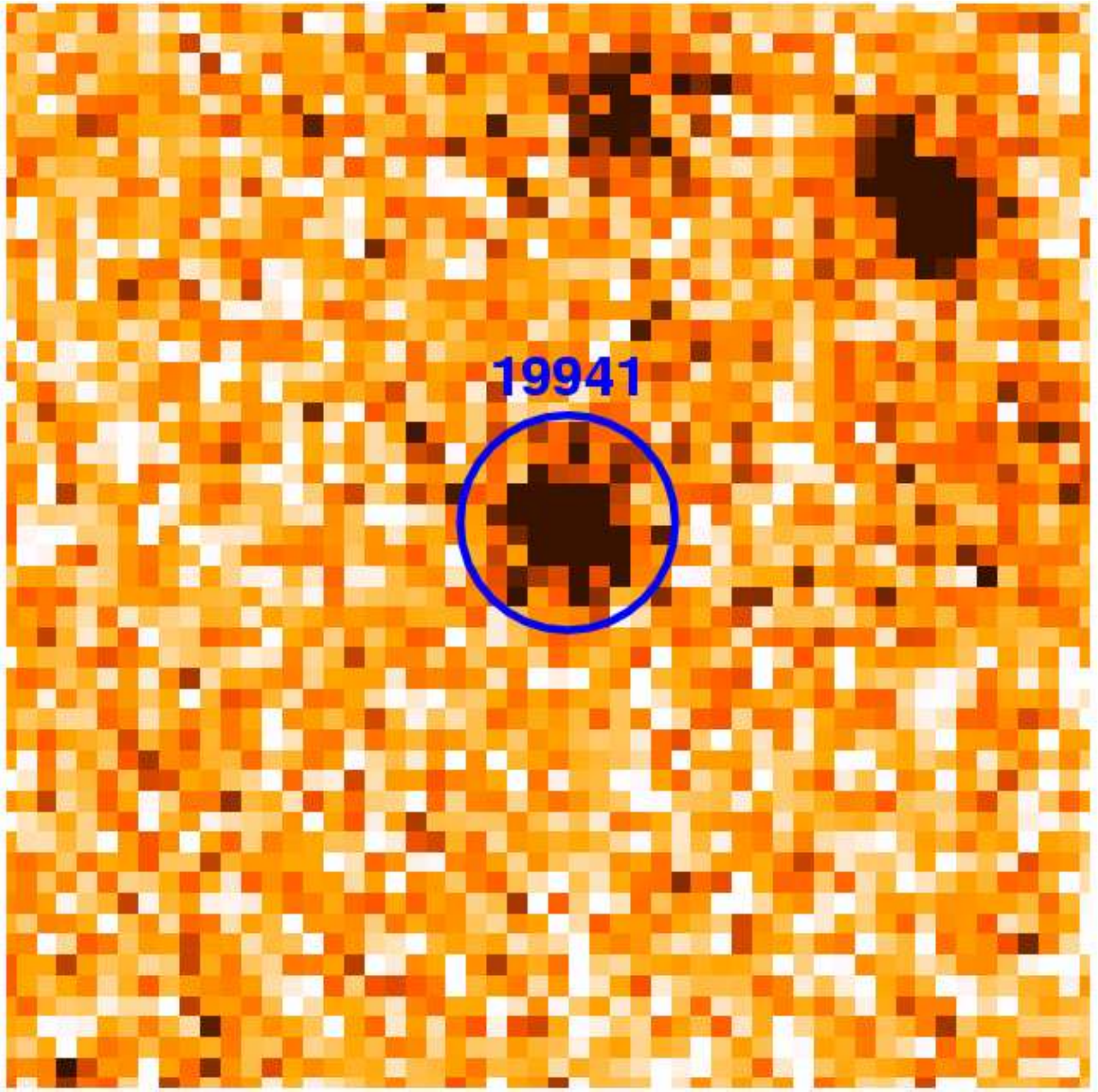}}}
\scalebox{0.21}[0.21]{\rotatebox{0}{\includegraphics{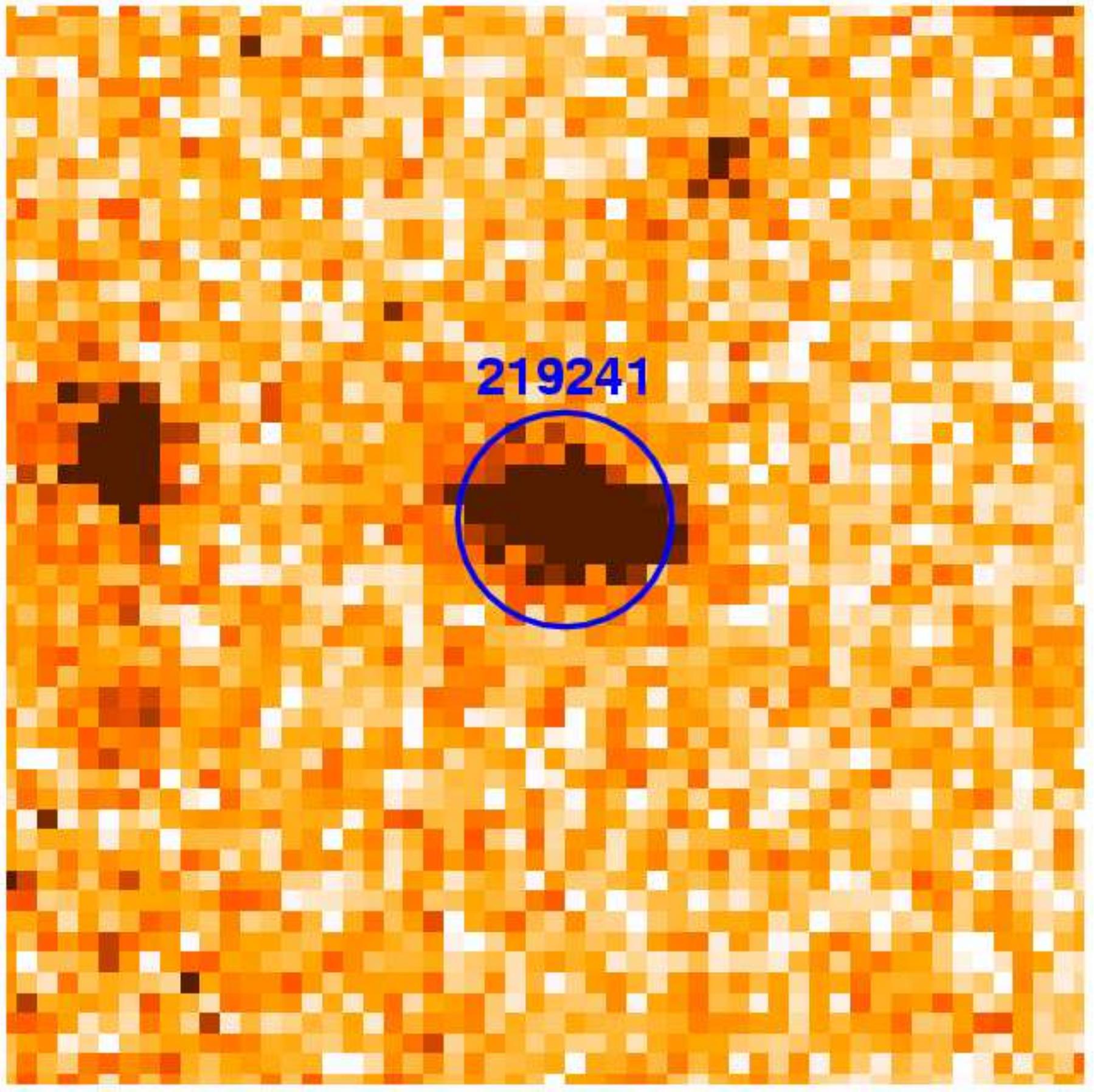}}}
\scalebox{0.21}[0.21]{\rotatebox{0}{\includegraphics{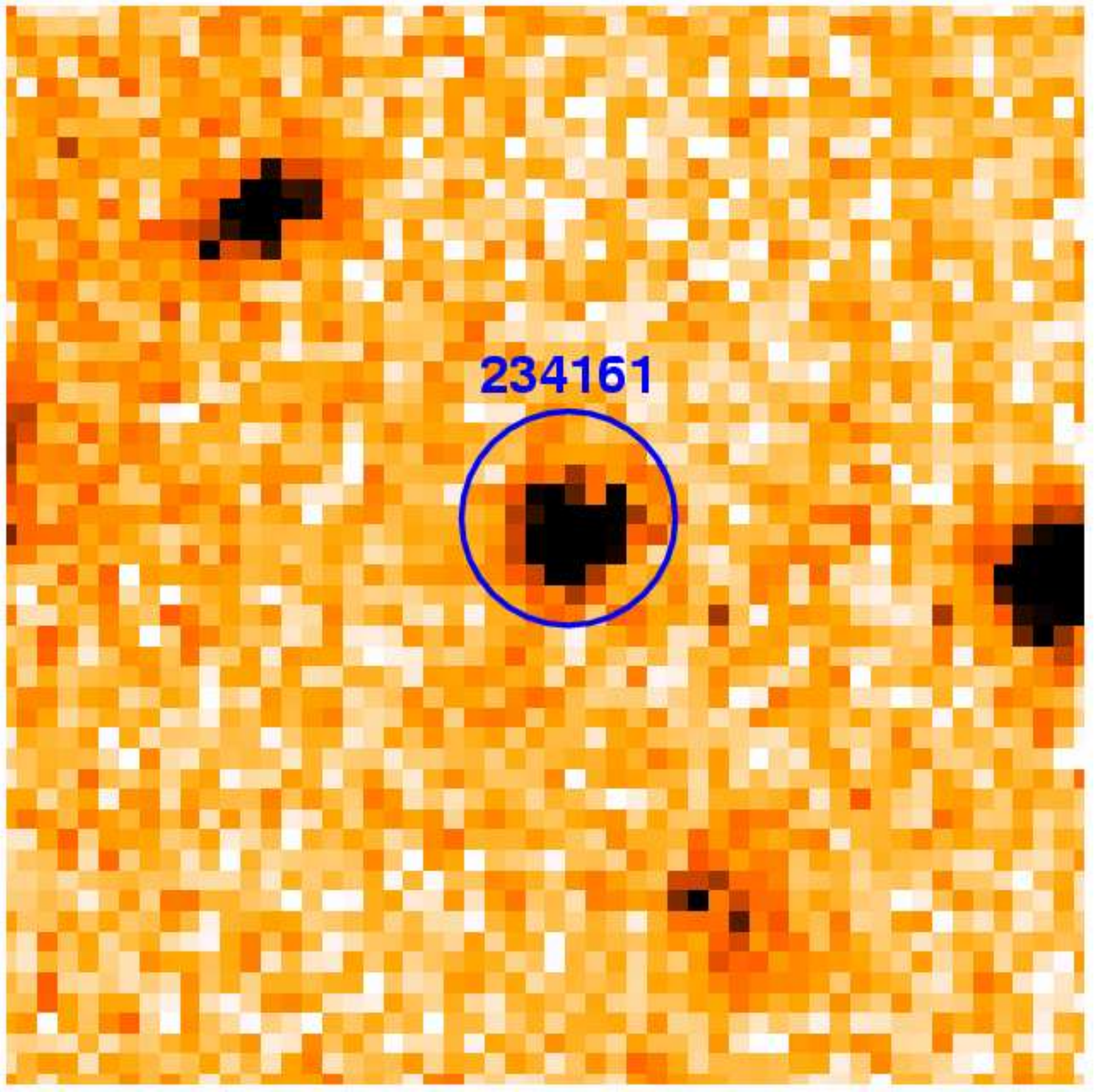}}}
\scalebox{0.21}[0.21]{\rotatebox{0}{\includegraphics{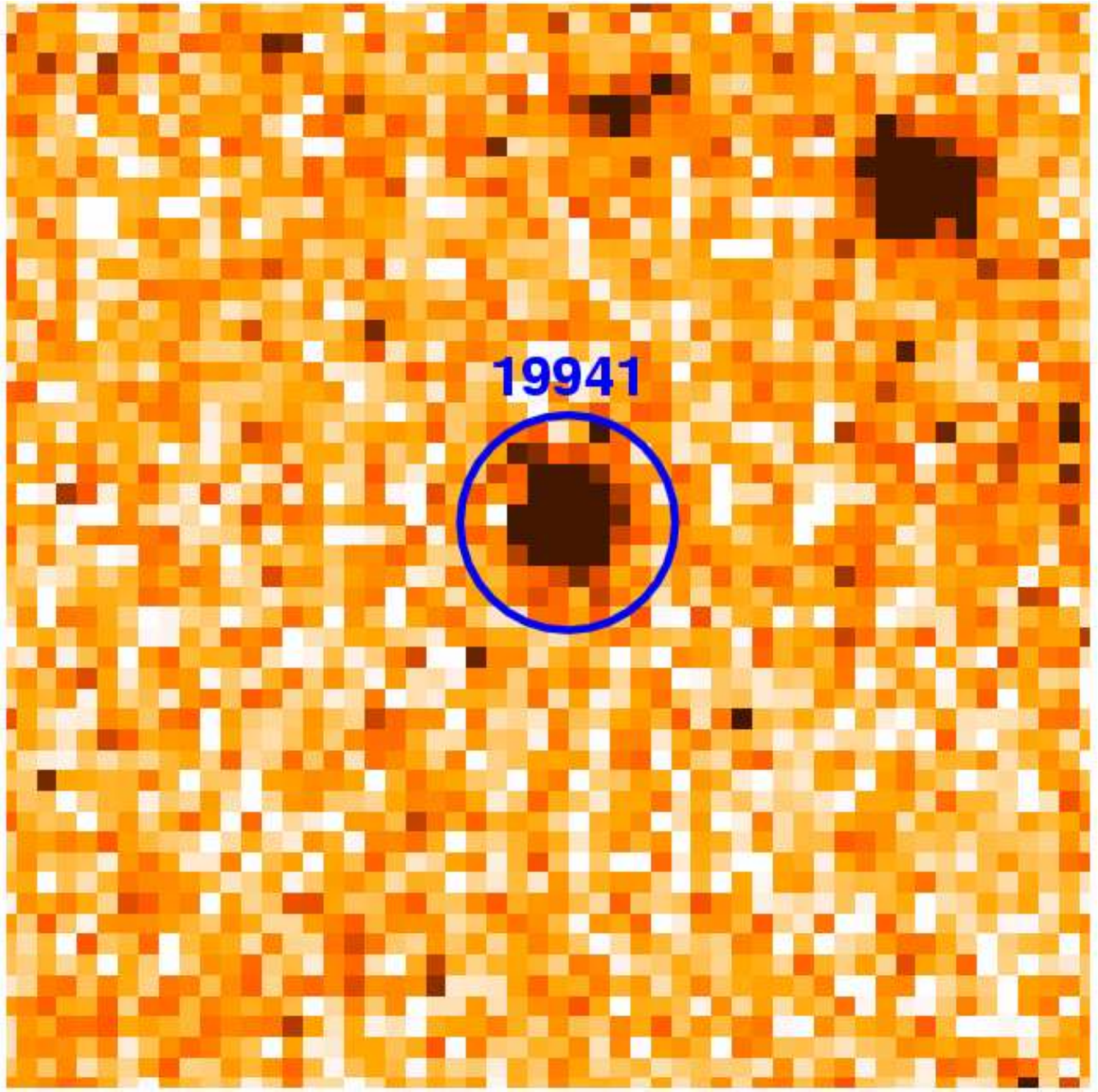}}}
\scalebox{0.21}[0.21]{\rotatebox{0}{\includegraphics{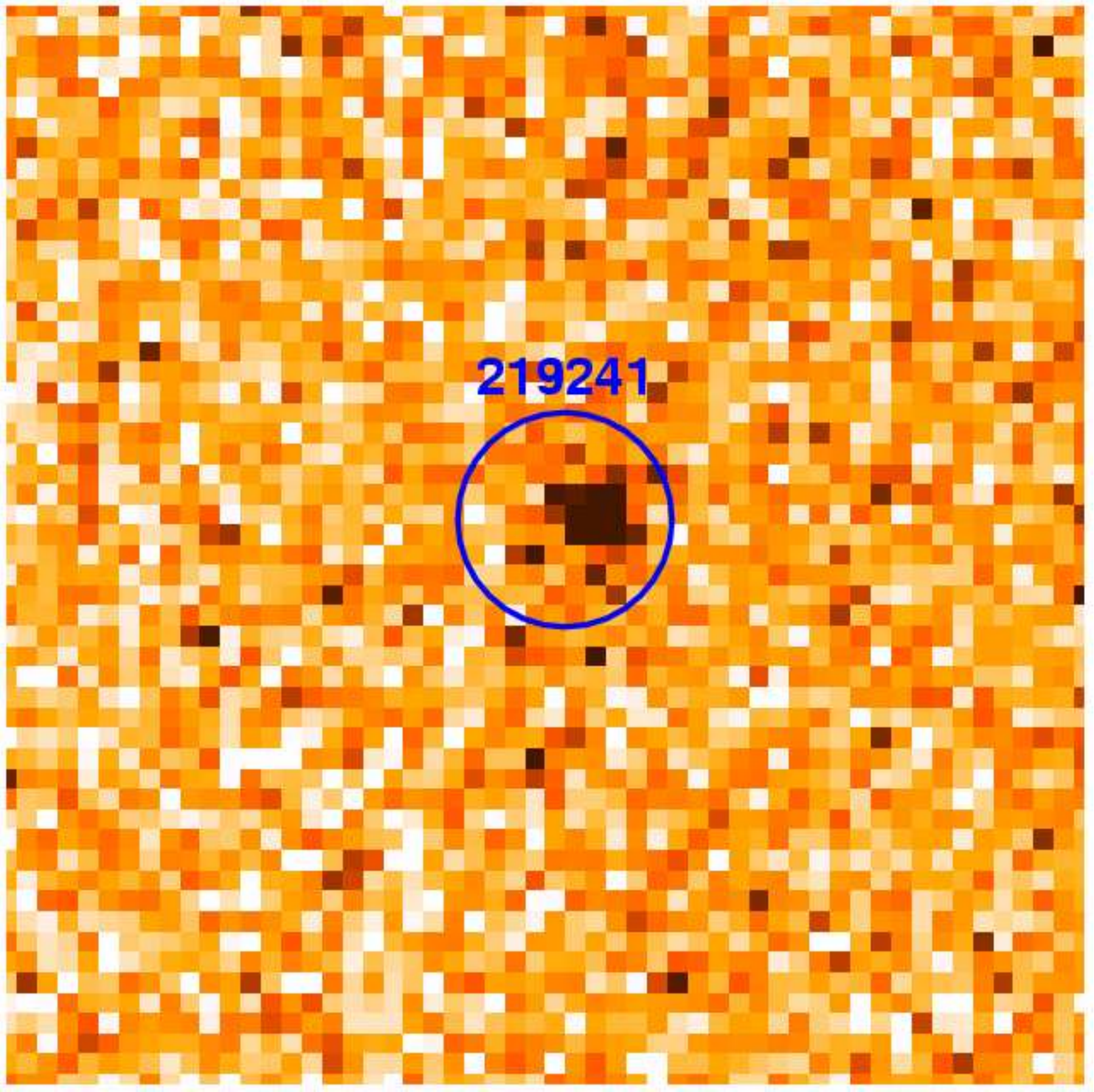}}}
\scalebox{0.21}[0.21]{\rotatebox{0}{\includegraphics{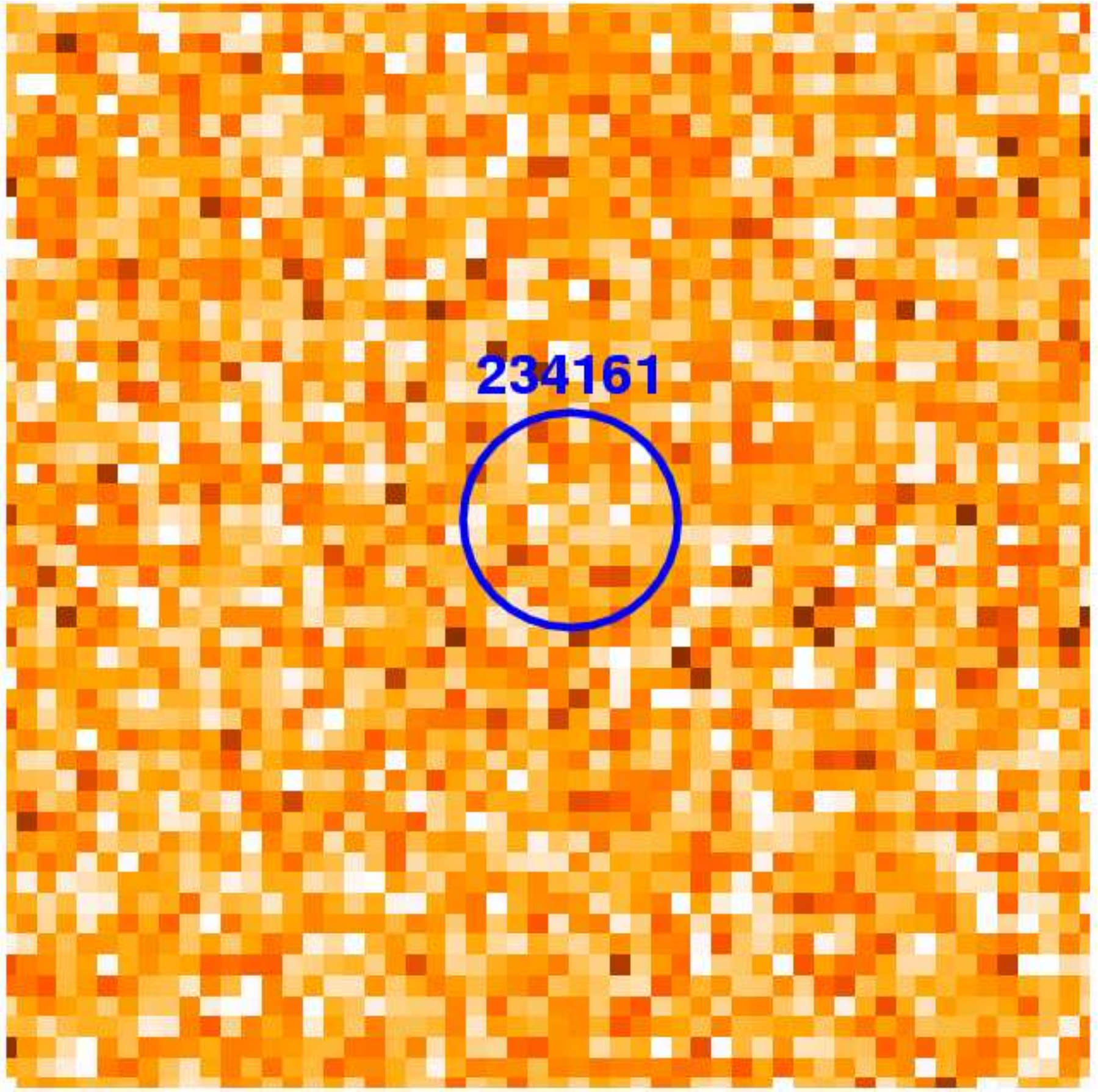}}}
\scalebox{0.21}[0.21]{\rotatebox{0}{\includegraphics{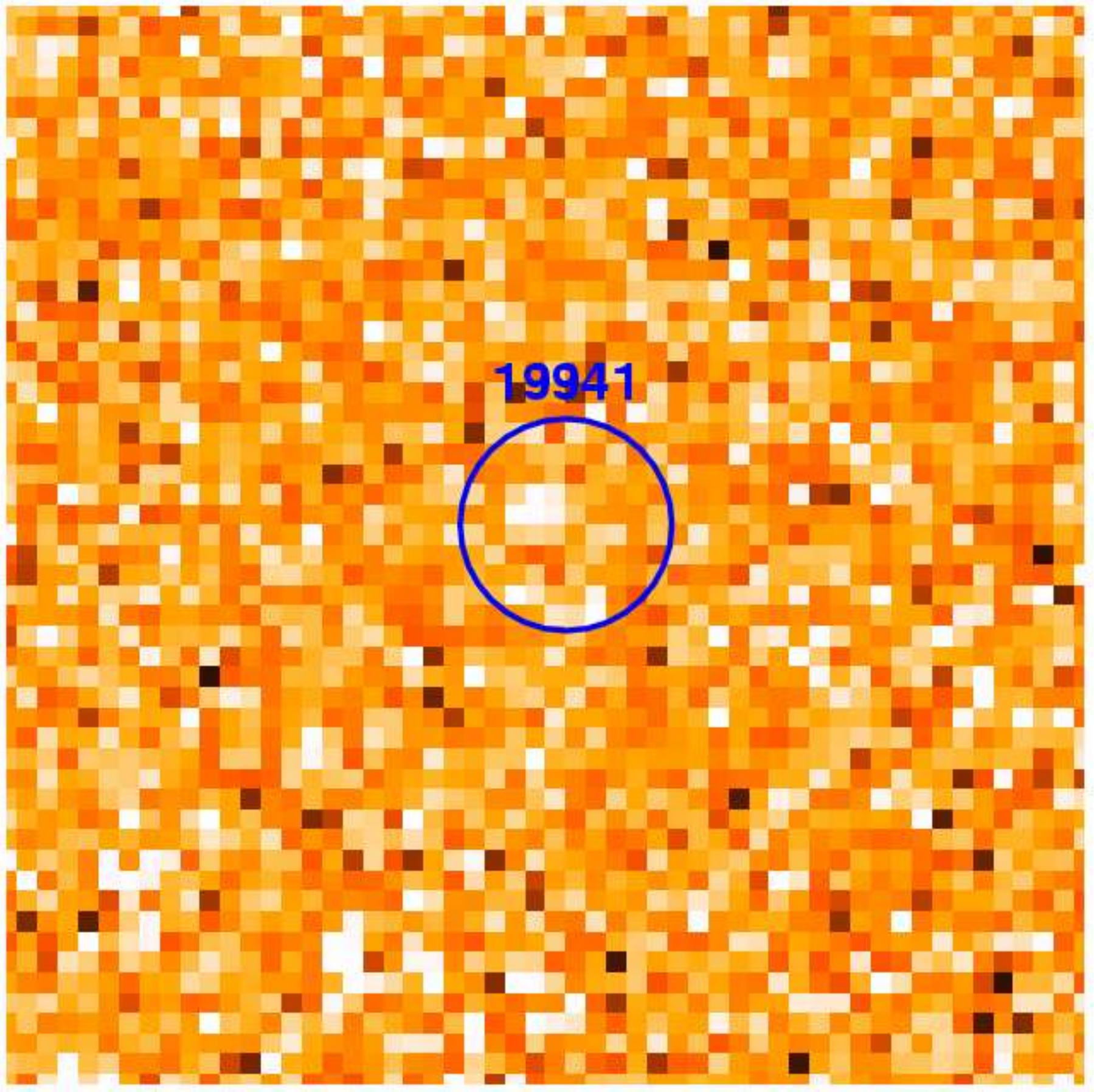}}}
\scalebox{0.21}[0.21]{\rotatebox{0}{\includegraphics{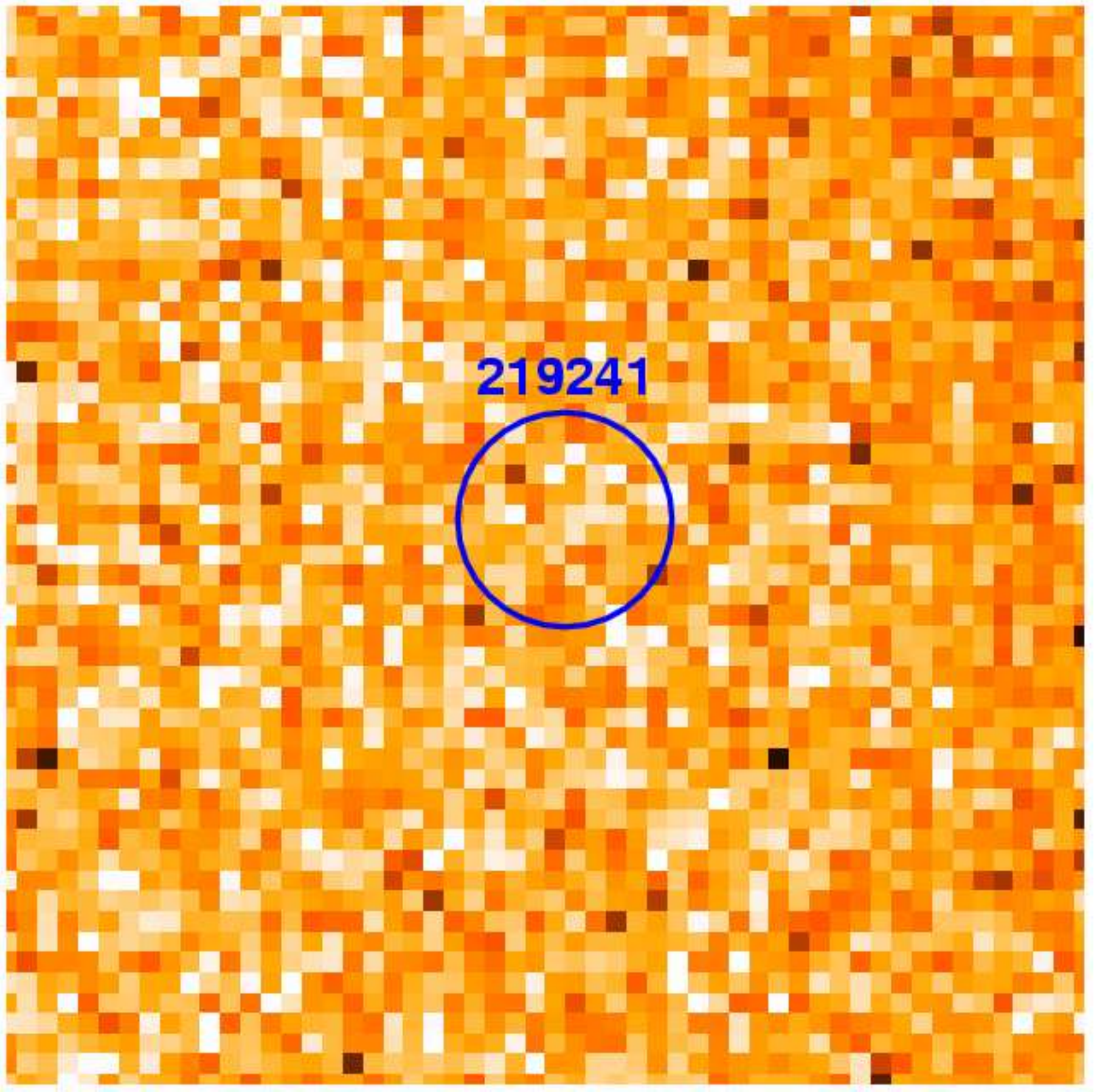}}}
\scalebox{0.21}[0.21]{\rotatebox{0}{\includegraphics{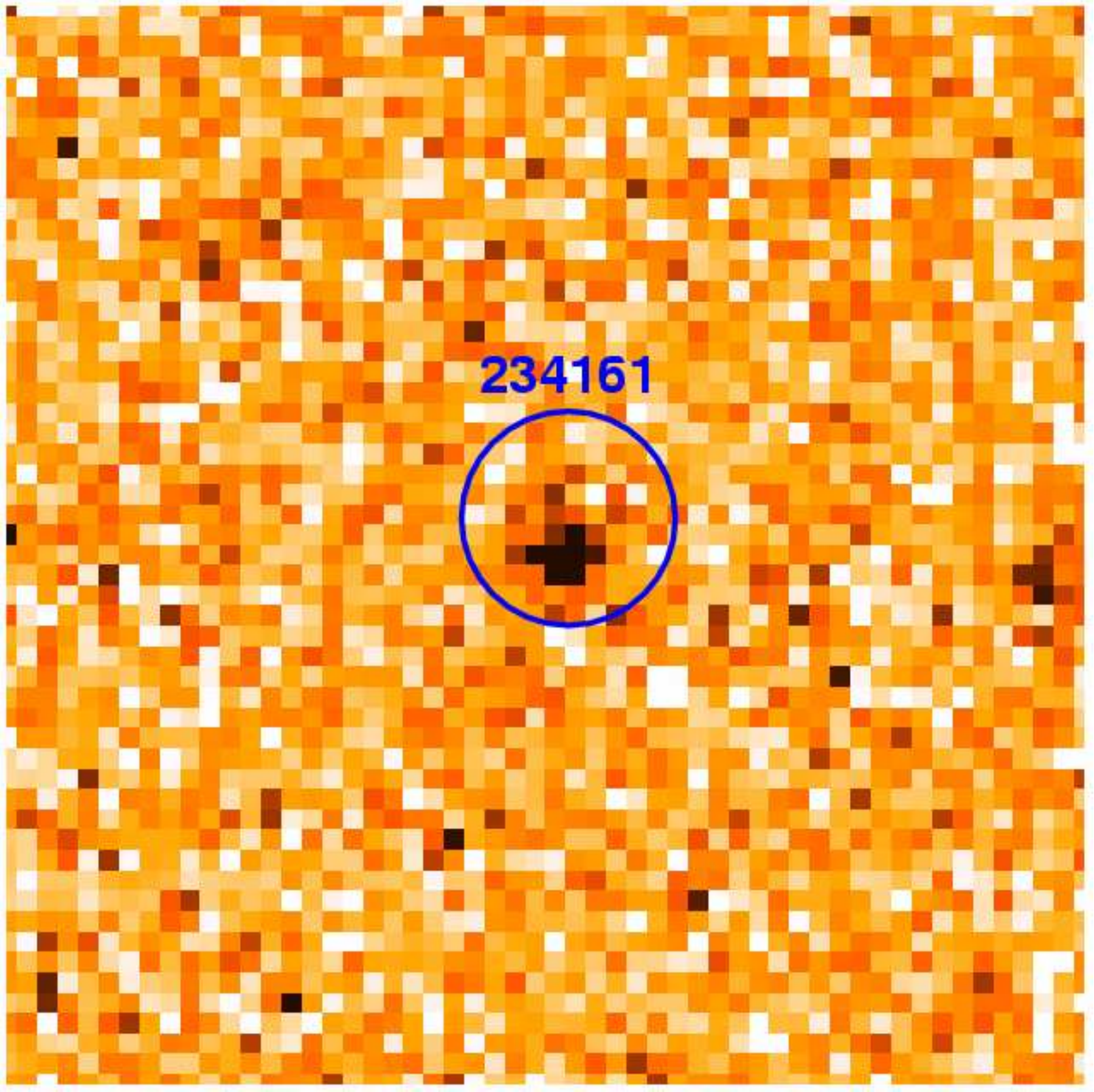}}}
\scalebox{0.21}[0.21]{\rotatebox{0}{\includegraphics{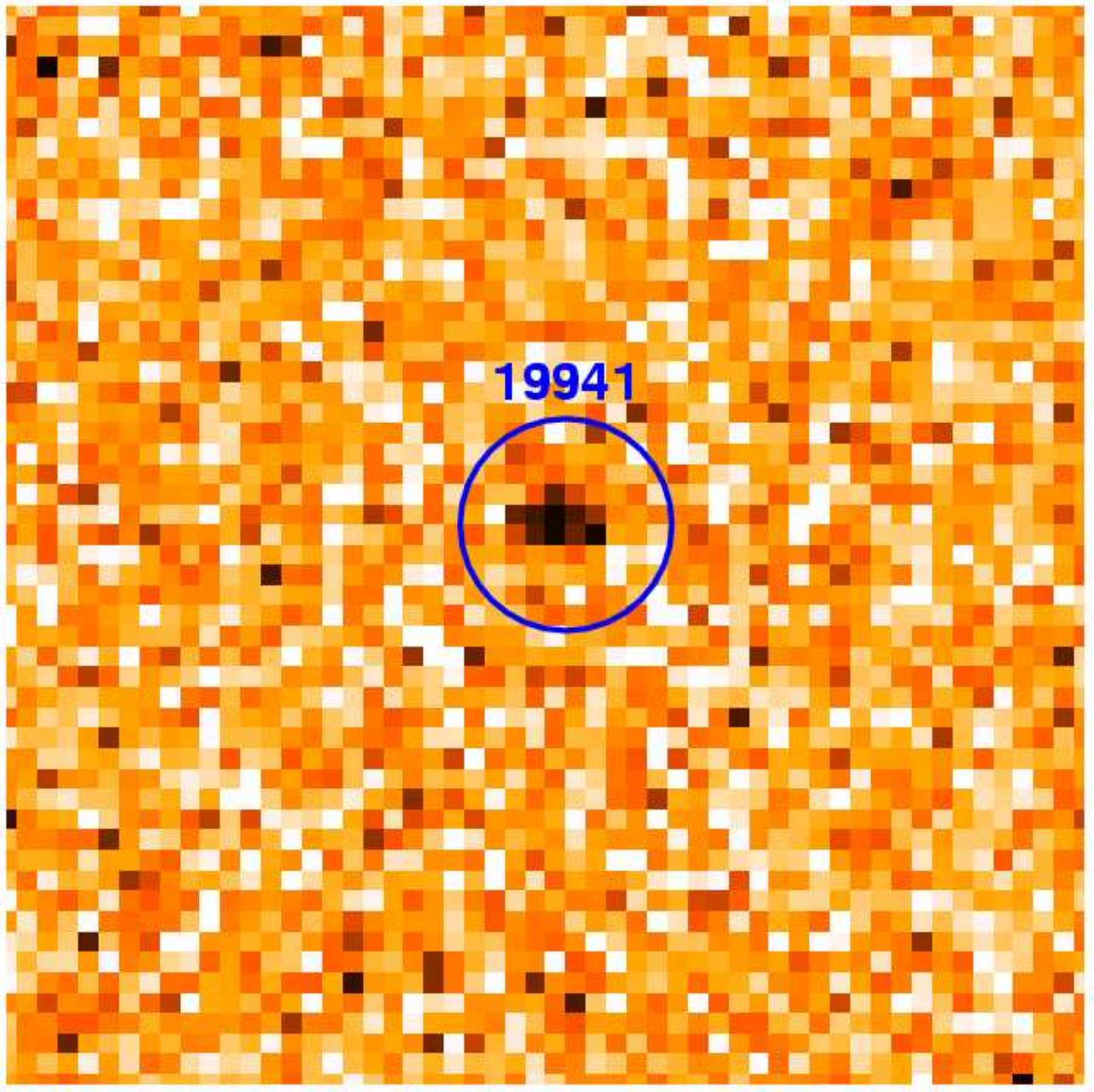}}}
\caption{Images and subtracted images in the $r'$-band for the three
  supernovae and their host galaxies.  Four $10''\times10''$ image
  sections are shown vertically for each supernova and are centred on
  their respective host galaxy.  Rows from top to bottom: Season 2004,
  season 2005, the subtracted image in 2004, and the subtracted image
  in 2005.  Columns from left to right, the images are of SN 219241,
  SN 234161, and SN 19941.  Each supernova exhibits a clean
  point-source detection in the subtracted image for each filter and a
  $\sim1-3$ kpc, physical, offset from its host galaxy centroid.
\label{rimages}}
\end{center}
\end{figure}

The SNe discussed in the main Letter are detected at the following
significance above their host galaxy flux in the $g', r',$ and $i'$
seasonal-stacked images, respectively, SN 219241: $6.4, 6.7, 8.4$; SN
234161: $3.0, 3.9, 4.2$; and SN 19941: $3.1, 5.1, 4.9$.  These are
determined using the SN integrated magnitudes in the seasonal-stacked
images and are $0.6-1.2$ magnitudes fainter in a given filter as
compared to the peak magnitudes determined from the light curves.
Figure~\ref{rimages} displays the $r'$-band images of the three SNe.
Each SN host galaxy is shown as it appeared in the 2004 and 2005
season and is indicated by a circle centred on its flux centroid.  The
SNe directly affect the appearance of the host galaxy in each case.
The subtracted images for both seasons are also shown.  The seasons
that do not include the SNe help to illustrate the clean subtraction
possible with the high-quality CFHT-LS images.  The SNe are
immediately obvious in the seasons with detections and have measurable
offsets from their host galaxy centroids.  The subtracted images are
similar in the $g'$ and $i'$-bands, as suggested by the detection
significance above, and the SN offsets from their host galaxy flux
centroids are consistent for each event filter to filter.

The seeing in the ``seasonal-stacked'' images is found to be
$\sim0.7''$ FWHM and stable across the images and over the four
epochs.  We subtracted the average flux of the ``seasonal-stacked''
images from the combined three adjacent years to construct the
subtracted images.  The three SNe exhibit clean radial profiles with
FWHMs equivalent to that of nearby point sources in the field.  The
pixel scale of the CFHT MegaCam is $0.187''$ pixel$^{-1}$ and
corresponds to $1.4-1.6$ kpc, physical, over $0.8<z<2.4$.  As listed
in Table 1 of the main Letter, the SNe have $1.1-2.8$ kpc separations
from their host galaxy centroids.  We tested the accuracy of these
separations by measuring the centroids of a galaxy sample that
includes the SN host galaxies and galaxies with similar magnitude.  We
determined that the centroid measurements are accurate to $\pm0.360$
pixels, $1\sigma$, in images tested season-to-season and $\pm0.175$
pixels, $1\sigma$, when tested filter-to-filter for a given season.
The SN centroids in the subtracted images were tested in an identical
manner and found to be accurate to $\pm0.493$ pixels, $1\sigma$,
filter-to-filter in their respective outburst season.  We conclude
that the SN-host centroid separations are accurate to $\pm0.523$
pixels, $1\sigma$.  This corresponds to $0.098''$ using the CFHT-LS
plate scale and $0.74-0.82$ kpc for the range of SN redshifts
presented here.

We performed an additional test of the separations between the
centroids of known AGN and their host galaxies.  These are AGN found
to exhibit a flux variation during the seasons probed by the CFHT-LS
data and have host galaxies $\sim1-2$ magnitudes brighter than the
host galaxies tested above.  We find the separations accurate to
$\pm0.152$ pixels, $1\sigma$.  Although we expect the centroid
uncertainties to behave as a function of magnitude, this test provides
a sense of the separation uncertainty for known AGN in host galaxies
with relatively similar magnitudes and demonstrates the quality of the
CFHT-LS imaging dataset.  As a result, the measured SN separations
from their host galaxy centroids are real, however SN 19941 is only
significant to $\sim1\sigma$.

\section{Light curves}
\label{lightcurves}

\hspace{1.0cm} The flux evolution for each of the three $z\sim2$ SNe
is presented in Figure 1 of the main Letter and the magnitude
information for the $g'r'i$ filters is listed in Table~\ref{SNmags}
below.  The uncertainties in the absolute magnitudes are dominated by
the photometry with a negligible contribution from the uncertainties
in redshift.  The strong detection and evolution of flux in the
$g'r'i'$ filters, that probe the rest-frame UV for these SNe, places
immediate constraints on their classification.  First, type Ia SNe
show very little flux shortward of $\sim300$ nm from efficient
scattering of UV photons~$^{29}$ and would be below the sensitivity of
the seasonal-stacked images at $z\ge0.7$.  This rules out a type Ia
SNe for these events.  Second, type Ib/c and type II SNe, other than
SNe IIn, have luminosities too low, and/or that rise and decline too
rapidly and remain essentially flat~$^{15}$ in the UV as compared to
that observed for SN 219241, SN 234161 and SN 19941.  Instead, the
peak UV magnitudes and evolution of are consistent with those of SNe
IIn.  We tested the intrinsic UV luminosity of all SN types and find
that $>90$\% of all $z\ge0.8$ detections in the CFHT-LS $g'$-band data
are expected to be SNe IIn.

Table~\ref{decay} lists the flux decay rates for each SN.  The
rest-frame bandpasses that correspond to the observed $g',~r',$ and
$i'$-band filters for each SN are arbitrarily indicated as UV1, UV2,
and UV3, respectively.  SNe IIn have varying decay rates in the
optical, but are essentially cooling blackbodies in the UV at early
times.  We use the expectations of a low extinction, cooling blackbody
guided by the UV continuum evolution of SN 1998S~$^{13}$ to generate a
model for the SNe IIn UV decay rates.  We compute the rest-frame
bandpass of the observations for each SN and the expected decay rate
considering the appropriate time dilation.  The relevant values of the
model are also listed in Table~\ref{decay}.  The SNe closely follow
the UV decay rate expectations of the model events at the determined
redshifts.  Moreover, these values are in close agreement to the
publicly available $Swift$ UVOT data of SN 2007pk and SN 2008am {\it
(http://heasarc.nasa.gov/docs/swift/sne/swift\_sn.html)}.

\section{Spectroscopy}
\label{spectra}

\hspace{1.0cm} We obtained low-resolution spectroscopy of SN 219241
and SN 234161 at the W. M. Keck Observatory (Keck) on 30 September
2008 and 01 October 2008.  We used the DEep Imaging Multi-Object
Spectrograph (DEIMOS)~$^{26}$ with the 600 line mm$^{-1}$ grating and
a multi-object spectroscopic wavelength coverage of $400-900$ nm.  We
acquired three 1200s exposures of SN 219241 and five 1200s exposures
of SN 234161.  The data were reduced using the UCB $spec2d$
pipeline\footnote{http://astro.berkeley.edu/~cooper/deep/spec2d/}.
Because DEIMOS is not sensitive below $\sim400$ nm and because bright
night sky emission lines begin to dominate the spectra longward of
$\sim800$ nm, the data were most effective in securing $z\le1$
redshifts of the faint host galaxies from strong rest-frame optical
features such as [O\textsc{ii}]$\lambda$ 372.7 nm and Balmer features
and $z\ge2.3$ redshifts from strong rest-frame UV features that
include prominent \lya and ISM absorption features.  Objects with
redshifts between $1<z<2$ fall in the ``redshift desert'' which is a
range of redshift where galaxies exhibit flat continua and no strong
identifying features in observed-frame optical wavelengths.
Nevertheless, the UV absorption features over these rest-frame
wavelengths have been well-studied and the characteristic flat
continua help to make $z\sim2$ galaxy colour-selection very
efficient~$^{22,}$\cite{adelberger04}.  In addition to the DEIMOS
data, we obtained low-resolution spectroscopy of SN 19941 on 25
January 2009 using the Low-Resolution Imaging Spectrometer
(LRIS)~$^{27,28}$ at Keck.  We used the 400 line mm$^{-1}$ grism
blazed at 340 nm and the 400 line mm$^{-1}$ grating blazed at 850 nm
to utilise both the blue and red arms of LRIS.  The excellent blue
sensitivity of LRIS resulted in $\sim320-900$ nm wavelength coverage.
The data were reduced using conventional methods that include IRAF and
IDL tasks and analysed with in-house code.

\begin{figure}
\begin{center}
\scalebox{0.30}[0.25]{\rotatebox{90}{\includegraphics{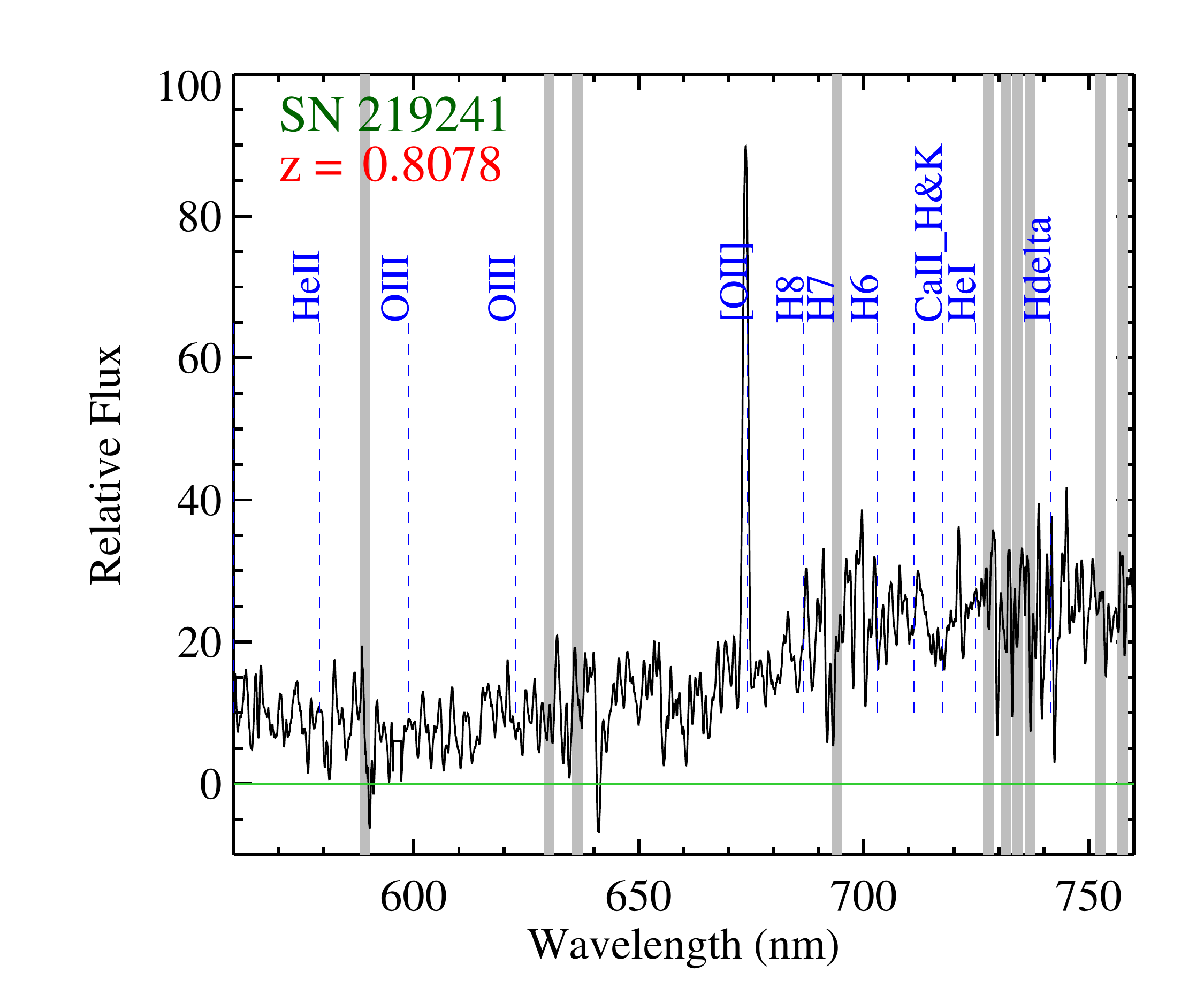}}}
\scalebox{0.30}[0.25]{\rotatebox{90}{\includegraphics{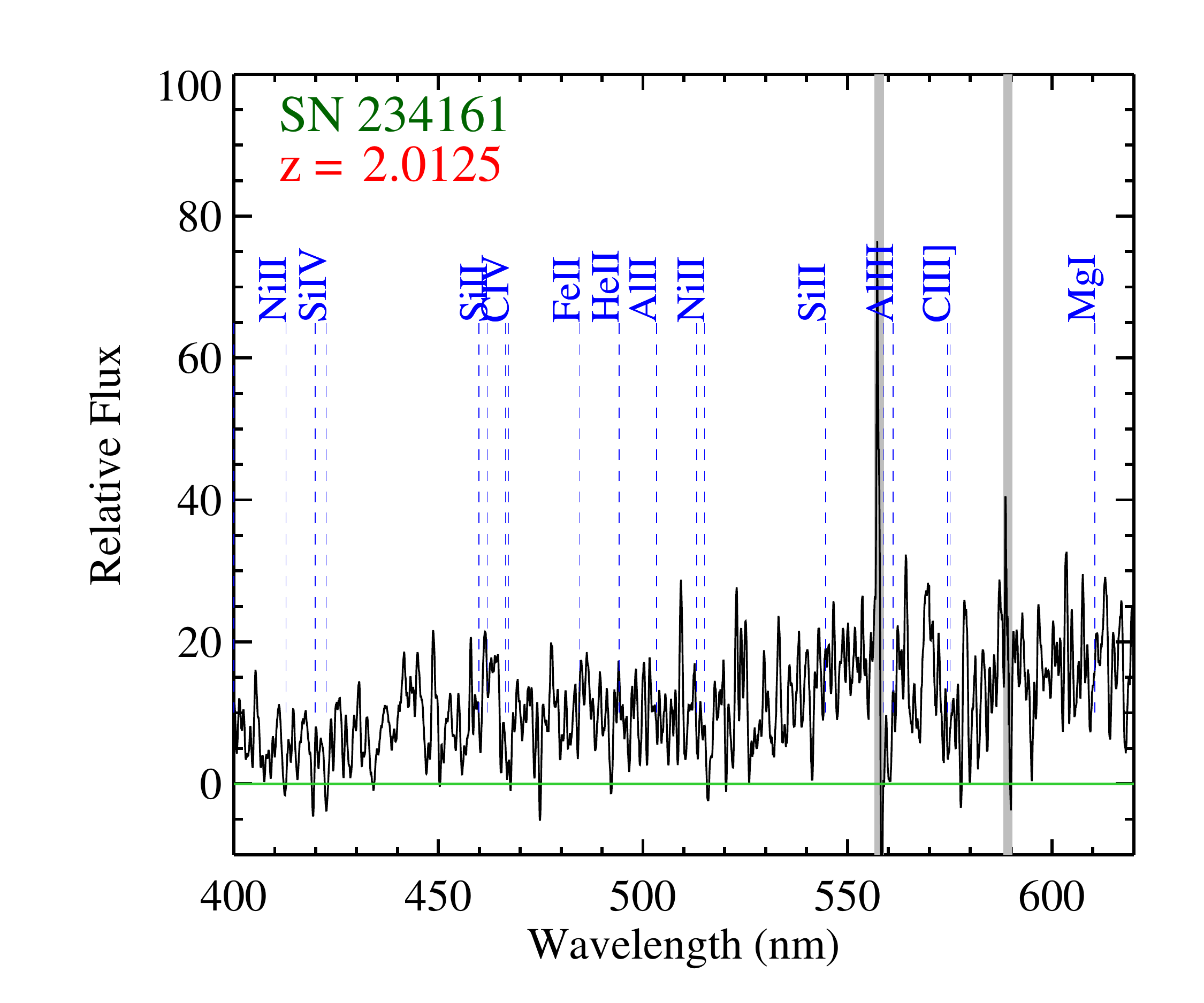}}}
\scalebox{0.30}[0.25]{\rotatebox{90}{\includegraphics{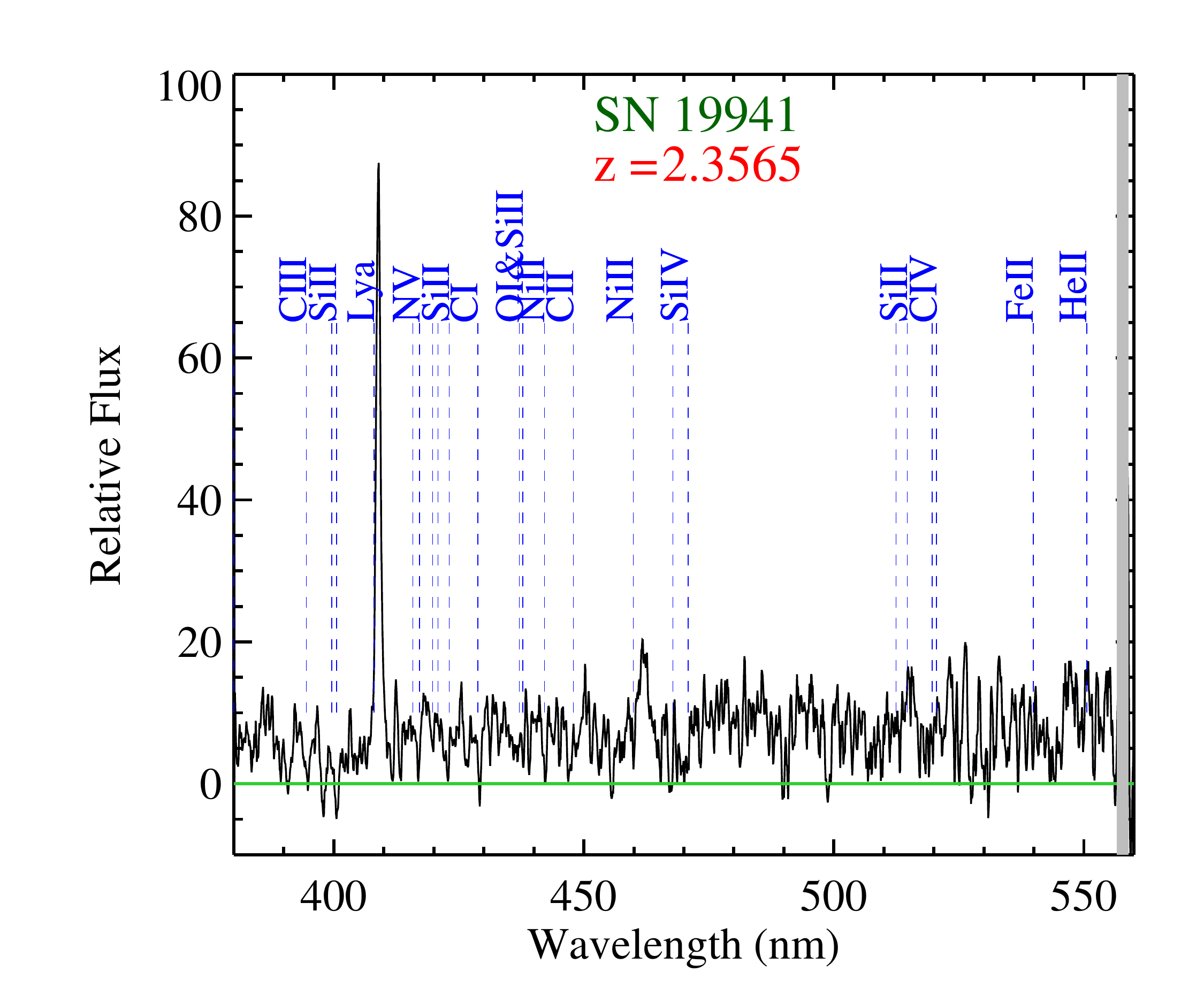}}}
\caption{\small Spectra of the high-redshift supernovae and host
  galaxies.  Thick vertical lines denote the positions of bright night
  sky emission lines that can be difficult to subtract cleanly from
  faint spectra.  The horizontal green line represents zero flux and
  the vertical dashed lines indicate expected transitions and are
  labelled.  {\it Top:} Strong [O\textsc{ii}]$\lambda$ 372.7 emission,
  Ca\textsc{ii}$\lambda\lambda$ 393.4, 396.8 absorption, and the break
  in the continuum near 400 nm are the dominant features that identify
  the $z=0.8078$ spectrum of SN 219241. {\it Centre:} The redshift of
  SN 234161 was determined by cross-correlation of $23$ of $30$ UV
  transitions that did not fall near bright night sky emission lines.
  We find a best fit redshift of $z=2.0125$ when considering the
  effect of the supernova emission lines (see Figure~\ref{lo}).  {\it
  Bottom:} The SN 19941 host galaxy exhibits strong identifying \lya
  emission.  The relatively strong blueshifted Si\textsc{iv}
  $\lambda\lambda$ 139.4, 140.3 nm and weaker C\textsc{iv}
  $\lambda\lambda$ 154.8, 155.0 nm and He\textsc{ii} 164.0 nm emission
  are tentatively assumed to originate from the supernova}
\label{1Dspectra}
\end{center}
\end{figure}

Figure~\ref{1Dspectra} presents the DEIMOS and LRIS spectra for the
three high-redshift SNe.  The $z=0.8078$ host galaxy of SN 219241
exhibits strong [O\textsc{ii}]$\lambda$ 372.7 emission and Ca H\&K
absorption lines.  SN 234161 host galaxy has a best fit redshift of
$z=2.0125$ from a cross-correlation of 23 ISM absorption features that
include Si\textsc{iv} $\lambda$ 139.4, 140.3, C\textsc{iv} $\lambda$
154.8, 155.1, Mg\textsc{i} $\lambda$ 202.6, 285.2, and Fe\textsc{ii}
$\lambda$234.4, 237.5, 238.3 after omitting those near bright night
sky emission lines.  SN 19941 exhibits strong \lya emission and is
found to have a redshift of $z=2.3565$ from a best fit to multiple ISM
absorption features.  Both SN 19941 and SN 234161 use a
prescription~\cite{adelberger03} to estimate the systemic redshift of
the host galaxy from the observed blueshifted ISM features and
redshifted \lya emission as a result of galactic winds.  The spectrum
of SN 19941 shows a relatively broad emission line near rest-frame
$\sim137$ nm ($\sim460$ nm observed-frame) and is discussed below.

\section{Detection of type IIn supernova emission}
\label{emission}

\hspace{1.0cm} In \S\ref{images} and \S\ref{lightcurves}, we showed
that the rest-frame UV photometry and light-curve evolution of the
three SNe best fit that of SNe IIn.  To confirm this, we search the
spectroscopy for evidence of the strong, long-lived emission lines
that define the IIn classification.  The emission lines are expected
to remain above the spectroscopic thresholds of 8m-class telescopes
using $\sim4$-hour exposures for $\sim2\cdot(1+z)$ years after
outburst~$^{14}$.  In addition, the emission lines peaks of SNe IIn
can have a $\sim1000-4000$ km s$^{-1}$
blueshift~$^{13}$\cite{filippenko89,leonard00} relative to the host
galaxy velocity and evolve with time.

\subsection{SN 219241 ---}
\label{219241}

This event was 855 days old, rest-frame, at the time observation.  At
this late stage, SNe IIn emission lines are expected to be near, or
below, the threshold of detection with a $\sim4$-hour observation.
Although the observing conditions were excellent ($0.4''$ seeing
FWHM), our combined 3600s exposure was likely insufficient for
significant emission-line detection.  As a result, we do not have
confident detection of SN line emission from SN 219241.  Deeper
optical or IR observations in the very near future may provide the
sensitivity to detect SN emission or to rule out a IIn classification.
As a result, the IIn classification is based on the UV photometry.

\subsection{SN 234161 ---}
\label{234161}

We searched for SN IIn emission line evidence in the combined 6000s
exposure obtained under same excellent seeing conditions as SN 219241.
This SN was observed 365 days old, rest-frame, and caught at a time
when SNe IIn emission lines are expected to be near maximum and above
the detection threshold.  We detect N\textsc{iv}] $\lambda\lambda$
146.1, 148.3, 148.8, C\textsc{iv} $\lambda\lambda$ 154.8, 155.0,
C\textsc{iii}] $\lambda\lambda$ 190.7, 190.9, and potentially
N\textsc{iii}] $\lambda\lambda$ 174.1, 174.7, 175.3 nm emission to
$>3-18\sigma$ significance (N\textsc{iii}] emission-line measurements
are less reliable because they fall near a subtracted weak night sky
emission line).  These emission lines arise from circumstellar
interaction and are found in low-redshift SNe IIn spectra~$^{12,13}$.

\begin{figure}
\begin{center}
\scalebox{0.47}[0.42]{\rotatebox{90}{\includegraphics{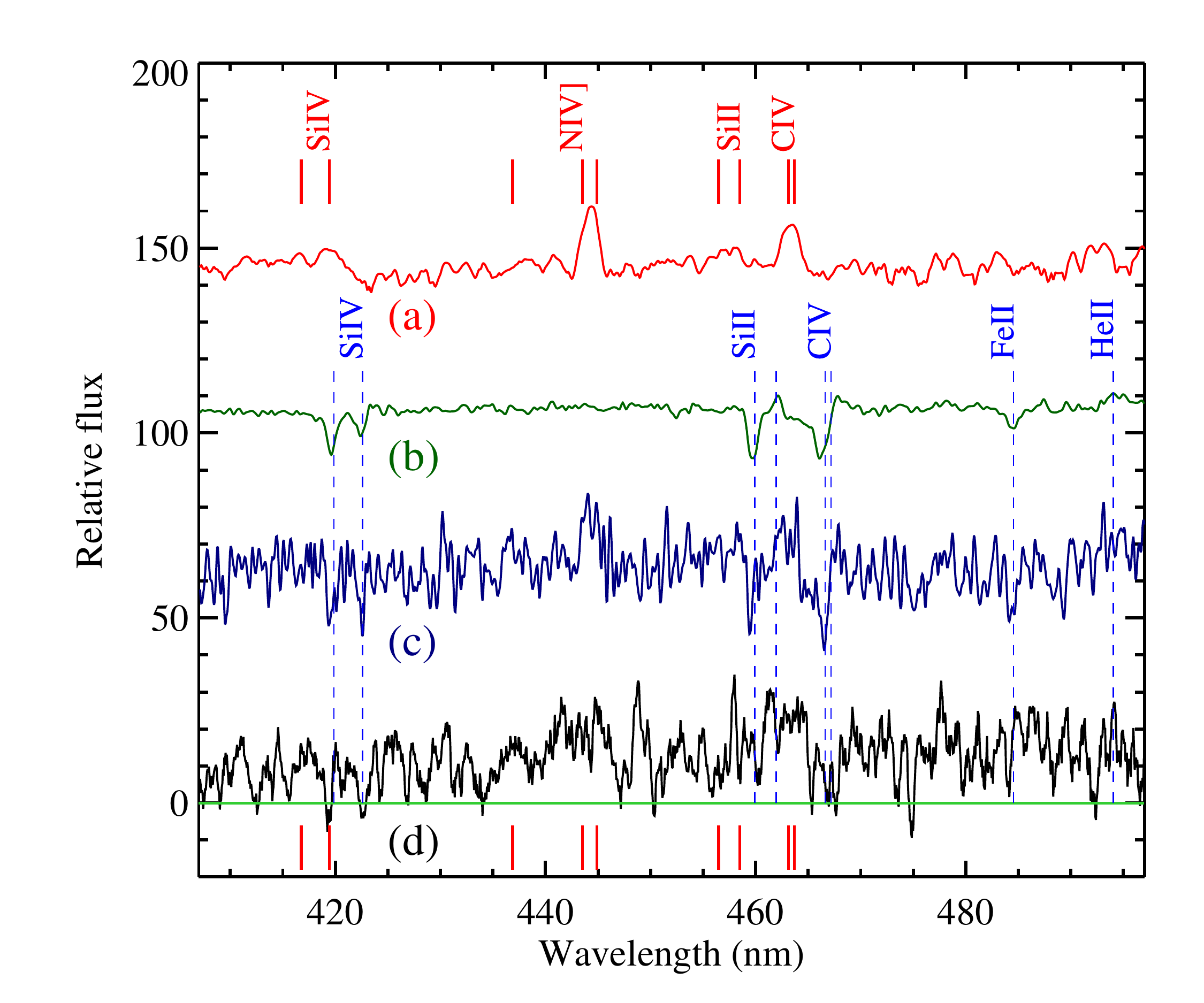}}}
\scalebox{0.47}[0.42]{\rotatebox{90}{\includegraphics{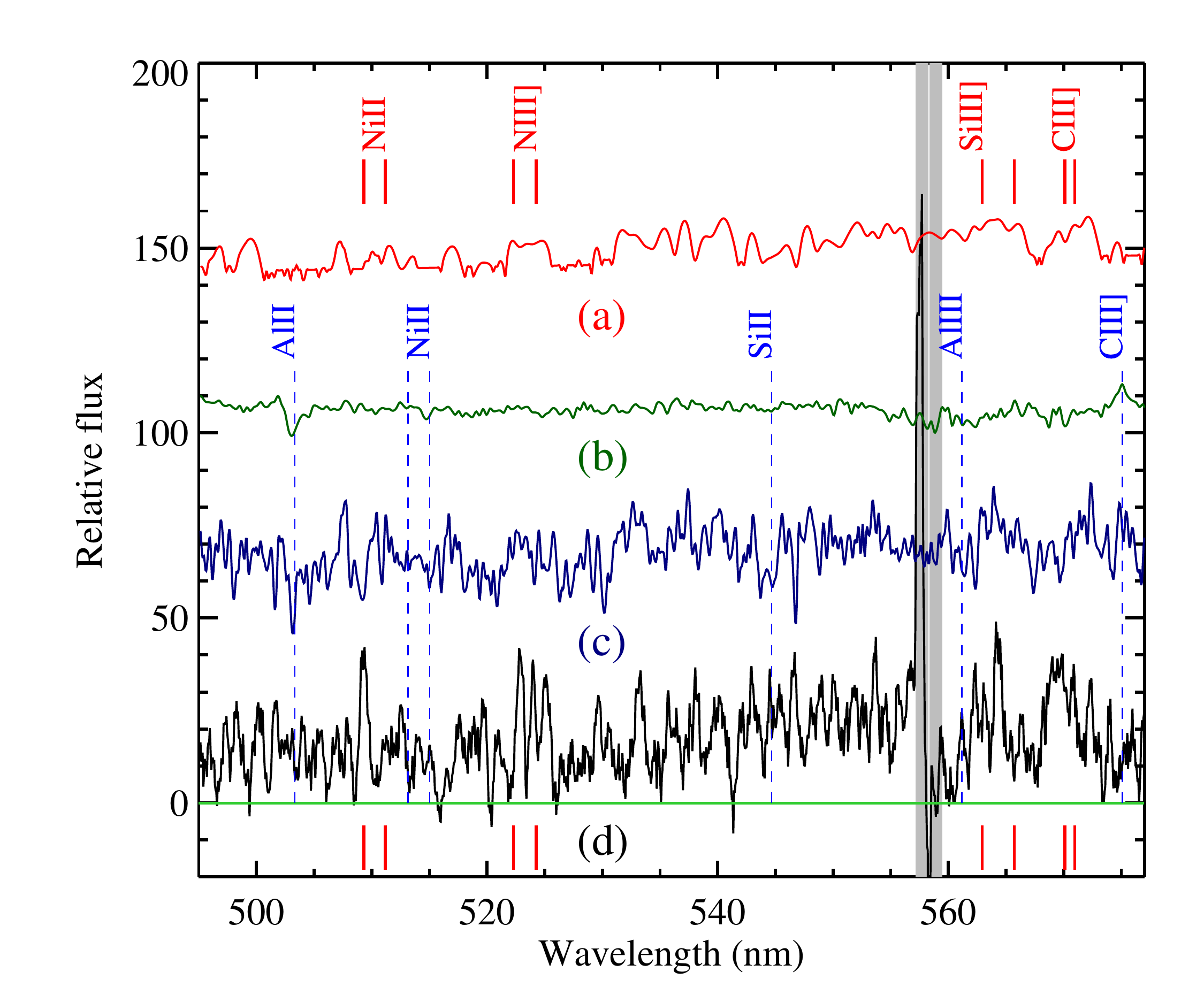}}}
\caption{\small Four spectra with arbitrary vertical offsets to aid in
  SN 234161 emission-line identification.  This plot is similar to
  Figure 2 in the main Letter but expanded into two wavelength
  sections for more detail.  The four spectra are: (a) the
  low-redshift SN 1998S~$^{13}$ (aged 485d) corrected to the redshift
  and observed velocity offset of SN 234161 ($z=2.0125-2500$ km
  s$^{-1}$), (b) a $z\sim3$ galaxy composite spectrum~$^{30}$
  corrected to $z=2.0125$.  (c) convolution of the SN 1998S and galaxy
  composite spectra with Gaussian noise added to match that of the
  data, and (d) the spectrum of SN 234161 (aged 365d).  See text for
  details.}
\label{lo}
\end{center}
\end{figure}

Figure~\ref{lo} is similar to Figure 2 in the main Letter but presents
the spectrum of SN 234161 separated in bluer and redder wavelength
regions for more detailed inspection.  Because the signal-to-noise
ratio (SNR) of the spectrum is only a few (host galaxy m$_{r'}=24.9$)
and the emission lines from the SN at such a high redshift are faint,
we present a high-redshift galaxy composite spectrum and the spectrum
of SN 1998S as an aid.  Figure~\ref{lo} steps through the spectra by
presenting from top to bottom: (a) the spectrum of SN 1998S~$^{13}$ at
day 485 shifted and flux corrected to $z=2.0125$ and blueshifted by
$2500$ km s$^{-1}$ to match the best-fit to the SN emission observed
in SN 234161, (b) a composite spectrum of 800 $z\sim3$ Lyman break
galaxies (LBGs)~$^{30}$ corrected to $z=2.0125$ that is representative
of $z\sim2-3$ galaxies~$^{21,22}$ color-selected via the technique
presented here, (c) a convolution of the composite LBG and SN 1998S
spectra with Gaussian noise added to match that of the data, followed
by (d) the spectrum of SN 234161.  Comparison of the feature profiles
and overall continuum shape of the data to the convolved spectrum show
the data is consistent with our interpretation of a $z=2.0125$ galaxy
hosting a SNe IIn event with $-2500$ km s$^{-1}$ offset emission.  We
use the spectrum of the type IIn(/L) SN 1998S here because it is the
only high-quality multiply-sampled UV SNe IIn spectrum to date.  We
use the older (day 943) single observation of the UV spectrum of SN
1995N~$^{12}$ throughout our analysis as well which exhibits similar
emission lines and relative line-strengths and ratios as those seen in
SN 234161.

Some of the effects of the SN can be seen in (1) the continuum profile
near 420 nm and the change in the Si\textsc{iv} doublet line strength
appearance in both the convolved spectrum and the data as a result of
SN Si\textsc{iv} emission, (2) the excess emission on a flat
featureless section of the galaxy spectrum near 440 nm attributed to
SN N\textsc{iv}] multiplet emission, (3) the continuum profile and
excess flux near 460 nm attributed to SN C\textsc{iv} emission and
host galaxy Si\textsc{ii} and the C\textsc{iv} absorption features,
(4) excess flux near 525 nm attributed to SN N\textsc{iii}] emission,
and (5) the clean detection of C\textsc{iii}] and weaker
Si\textsc{iii]} emission from the SN near 570 nm in the data.  The
spectrum of SN 1998S shown here is 120 days older than SN 234161.  SN
1998S exhibited stronger N\textsc{iii}] and C\textsc{iii}] emission at
earlier times~$^{13}$ as did SN 1995N at a later time~$^{12}$.

\subsection{SN 19941 ---}
\label{19941}

This SN was 345 days old, rest-frame, at the time of the LRIS
observations.  The salient feature in the host galaxy spectrum is the
strong \lya emission near observed wavelength 410 nm.
Cross-correlation with ISM absorption features results in a best fit
redshift of $z=2.3565$.  This is the highest redshift SN ever
detected.  SN 19941 exhibited a flux increase in only one season and
was stable in three, mimicking expected SN behavior, and shows no
evidence of AGN emission lines.  As with SN 234161 above, this SN is
in the time regime when the emission lines from the circumstellar
interaction are expected to be near their strongest.  This spectrum
shows a strong, relatively broad emission line near $\sim460$ nm
($\sim137$ nm rest-frame) that may be the result of blueshifted
Si\textsc{iv} $\lambda\lambda$139.4, 140.3 nm emission from the SN.
We also find excess C\textsc{iv} and He\textsc{ii} 164.0 nm flux near
observed 515 and 550 nm, respectively, which would be in accordance
with this interpretation.  The SNR of the reduced spectrum obtained
with the less sensitive red CCD of LRIS was too low to make any
significant assessment other than the lack of strong emission redward
of $\sim560$ nm $(\sim170$ nm, rest-frame).  This event will require
further analysis and deep, high-SNR spectroscopy to search for line
evolution.

Over the UV wavelength range probed in this work, the main features
that distinguish the SN detections from AGN are the N\textsc{iv]},
N\textsc{iii]} and C\textsc{iii]} strengths and their ratios with
C\textsc{iv}.  SN 234161 exhibits $\sim1:1$ C\textsc{iv}/N\textsc{iv]}
ratio and stronger C\textsc{iii]} than C\textsc{iv}, with
C\textsc{iv}/C\textsc{iii]} $\sim0.3$.  These strengths are similar to
that seen in the two available late-time SNe IIn UV spectra; namely
that of SN 1995N and SN 1998S~$^{12,13}$.  Over similar epochs, SN
1995N and SN 1998S exhibit C\textsc{iv}/N\textsc{iv]} ratios of
$\sim0.6-3$ and C\textsc{iv}/C\textsc{iii]} ratios of $0.3-1$.  In
contrast, the SDSS QSO composite spectrum~\cite{vandenberk01} shows no
evidence of N\textsc{iv]} emission and has a
C\textsc{iv}/N\textsc{iii]} ratio of $\sim66$.

Both N\textsc{iv]} and N\textsc{iii]} are indicators of shock
ionisation and only about a dozen known objects at high redshift
(including AGN, radio galaxies, QSOs, and normal galaxies) exhibit
these lines in emission at a significant level.  Rare AGN that do show
these features, average a C\textsc{iv}/N\textsc{iv]} ratio of
$\sim20-25$, compared to $\sim1$ for SN 234161.  These rare AGN still
maintain C\textsc{iv}/C\textsc{iii]} ratios of $1.5-2.0$ as is the
case for normal AGN (cf., C\textsc{iv}/C\textsc{iii]} $\sim0.3$ for SN
234161).  Of all objects at high redshift, including normal galaxies,
we are aware of only one object in the literature, DLS
1053+0528~\cite{glikman07} that shows equivalent or stronger
N\textsc{iv]} emission as compared to C\textsc{iv}.  This object does
not show other emission lines indicative of shock ionisation as those
seen in SNe IIn (e.g., N\textsc{iii]}).  C\textsc{iii]} information is
not presented for this object.  In summary, the UV emission lines seen
in the host galaxy spectra of SN 234161 and SN 19941 are unlike AGN
and are seen with the expected strengths, line ratios, and blueshifted
emission peaks reminiscent of SNe IIn.

\clearpage

\begin{table}
\centering
{\footnotesize
\begin{tabular}{lccccccc}
\hline
SN & Redshift & m$_{g'}$& m$_{r'}$& m$_{i'}$& $M_{UV1}$& $M_{UV2}$& $M_{UV3}$\\
   &    $z$      & peak    & peak    & peak    & peak     & peak     & peak  \\
\hline
219241 & $0.8078\pm0.0010$ & $25.14^{+0.15}_{-0.17}$ &
  $24.73^{+0.11}_{-0.12}$ & $24.35^{+0.12}_{-0.13}$ &
  $-17.75^{+0.15}_{-0.17}$ & $-18.16^{+0.11}_{-0.12}$ &
  $-18.54^{+0.12}_{-0.13}$ \\
234161 & $2.0125\pm0.0021$ & $25.65^{+0.25}_{-0.32}$ &
  $25.17^{+0.19}_{-0.24}$ & $25.07^{+0.26}_{-0.34}$ &
  $-19.12^{+0.25}_{-0.32}$ & $-19.61^{+0.19}_{-0.24}$ &
  $-19.71^{+0.26}_{-0.34}$ \\
 19941 & $2.3565\pm0.0022$ & $25.61^{+0.21}_{-0.26}$ &
  $25.59^{+0.20}_{-0.25}$ & $25.40^{+0.34}_{-0.49}$ &
  $-19.46^{+0.21}_{-0.26}$ & $-19.49^{+0.20}_{-0.25}$ &
  $-19.68^{+0.34}_{-0.49}$ \\
\hline
\hline
\end{tabular}
\caption{Redshift and magnitude information for the three
high-redshift SNe.  Magnitudes are at or near maximum brightness as
determined from the light curves.  Absolute UV magnitudes are
estimated for the effective rest-frame wavelengths probed by the
$g'r'i'$ filters.  UV1, UV2, and UV3 correspond to the $g',r'$ and
$i'$-bands, respectively.}
\label{SNmags}}
\end{table}

\clearpage

\begin{table}
\centering
{\footnotesize
\begin{tabular}{lcccccc} 
\hline 
SN& UV1 ($g'$)& UV1 Decay& UV2 ($r'$)& UV2 Decay& UV3 ($i'$)& UV3 Decay\\ 
 & nm & mag day$^{-1}$ & nm & mag day$^{-1}$ & nm & mag day$^{-1}$\\
\hline
219241 ($z=0.8078$)& $265$ & $0.079$ & $350$ & $0.045$ & $420$ &$0.044$\\
         IIn model & $270$ & $0.096$ & $370$ & $0.052$ & $420$ &$0.042$\\
234161 ($z=2.0125$)& $160$ & $0.175$ & $210$ & $0.102$ & $250$ &$0.064$\\
         IIn model & $170$ & $0.159$ & $220$ & $0.105$ & $270$ &$0.098$\\
 19941 ($z=2.3565$)& $145$ & $0.205$ & $190$ & $0.129$ & $225$ &$0.119$\\
         IIn model & $125$ & $0.204$ & $170$ & $0.152$ & $220$ &$0.112$\\
\hline
\hline
\end{tabular}
\caption{Supernova decay rates.  Listed are the effective UV
wavelengths probed by the $g, r',$ and $i'$-band filters and estimated
decay rates taken from the supernova light curves over $\sim15-35$
days, rest-frame, following peak or near peak brightness.  Below each
observed value is the expected value from the model.  Similar to
Figure~\ref{SNmags}, UV1, UV2, and UV3 correspond to the rest-frame
wavelengths probed by the $g',r'$ and $i'$-bands, respectively.  See
text for details.}
\label{decay}}
\end{table}

\clearpage

\clearpage

\clearpage

\clearpage

\begin{thebibliography}{10}
\expandafter\ifx\csname url\endcsname\relax
  \def\url#1{\texttt{#1}}\fi
\expandafter\ifx\csname urlprefix\endcsname\relax\def\urlprefix{URL }\fi
\providecommand{\bibinfo}[2]{#2}
\providecommand{\eprint}[2][]{\url{#2}}

\bibitem{poznanski07}
\bibinfo{author}{{Poznanski}, D.} \emph{et~al.}
\newblock \bibinfo{title}{{Supernovae in the Subaru Deep Field: an
    initial sample and Type Ia rate out to redshift 1.6}}
\newblock \emph{\bibinfo{journal}{Mon. Not. R. Astron. Soc.}}
\textbf{\bibinfo{volume}{382}}, 
  \bibinfo{pages}{1169--1186} (\bibinfo{year}{2007}).

\bibitem{riess01}
\bibinfo{author}{{Riess}, A. G.} \emph{et~al.}
\newblock \bibinfo{title}{{The Farthest Known Supernova: Support for an
    Accelerating Universe and a Glimpse of the Epoch of Deceleration}}
\newblock \emph{\bibinfo{journal}{Astrophysical Journal}}
\textbf{\bibinfo{volume}{560}}, 
  \bibinfo{pages}{49--71} (\bibinfo{year}{2001}).

\bibitem{botticella08}
\bibinfo{author}{{Botticella}, M. T.} \emph{et~al.}
\newblock \bibinfo{title}{{Supernova rates from the Southern inTermediate
      Redshift ESO Supernova Search (STRESS)}}
\newblock \emph{\bibinfo{journal}{Astronomy \& Astrophysics}}
\textbf{\bibinfo{volume}{479}},
  \bibinfo{pages}{49--68} (\bibinfo{year}{2008}).

\bibitem{dellavalle06}
\bibinfo{author}{{Della Valle}, M.} \emph{et~al.}
\newblock \bibinfo{title}{{Hypernova Signatures in the Late
    Rebrightening of GRB 050525A}}
\newblock \emph{\bibinfo{journal}{Astrophysical Journal}}
\textbf{\bibinfo{volume}{642}},
  \bibinfo{pages}{103--106} (\bibinfo{year}{2006}).

\bibitem{soderberg06}
\bibinfo{author}{{Soderberg}, A.~M.} \emph{et~al.}
\newblock \bibinfo{title}{{An HST Study of the Supernovae Accompanying 
    GRB 040924 and GRB 041006}}
\newblock \emph{\bibinfo{journal}{Astrophysical Journal}}
\textbf{\bibinfo{volume}{636}},
  \bibinfo{pages}{391--399} (\bibinfo{year}{2006}).

\bibitem{schlegel90}
\bibinfo{author}{{Schlegel} E. M.}
\newblock \bibinfo{title}{{A new subclass of Type II supernovae?}}
\newblock \emph{\bibinfo{journal}{Mon. Not. R. Astron. Soc.}}
\textbf{\bibinfo{volume}{244}}, 
  \bibinfo{pages}{269--271} (\bibinfo{year}{1990}).

\bibitem{richardson02}
\bibinfo{author}{{Richardson}, D.} \emph{et~al.}
\newblock \bibinfo{title}{{A Comparative Study of the Absolute Magnitude
    Distributions of Supernovae}}
\newblock \emph{\bibinfo{journal}{Astronomical Journal}}
\textbf{\bibinfo{volume}{123}}, 
  \bibinfo{pages}{745--752} (\bibinfo{year}{2002}).

\bibitem{smith07}
\bibinfo{author}{{Smith}, N.} \emph{et~al.}
\newblock \bibinfo{title}{{SN 2006gy: Discovery of the Most Luminous 
    Supernova Ever Recorded, Powered by the Death of an Extremely Massive
    Star like $\eta$ Carinae}}
\newblock \emph{\bibinfo{journal}{Astrophysical Journal}}
\textbf{\bibinfo{volume}{666}}, 
  \bibinfo{pages}{1116-1128} (\bibinfo{year}{2007}).

\bibitem{smith08}
\bibinfo{author}{{Smith}, N.} \emph{et~al.}
\newblock \bibinfo{title}{{SN 2006tf: Precursor Eruptions and the Optically
    Thick Regime of Extremely Luminous Type IIn Supernovae}}
\newblock \emph{\bibinfo{journal}{Astrophysical Journal}}
\textbf{\bibinfo{volume}{686}}, 
  \bibinfo{pages}{467-484} (\bibinfo{year}{2008}).

\bibitem{kotak06}
\bibinfo{author}{{Kotak}, R.} {\& Vink, J. S.}
\newblock \bibinfo{title}{{Luminous blue variables as the progenitors of
  supernovae with quasi-periodic radio modulations}}
\newblock \emph{\bibinfo{journal}{Astronomy \& Astrophysics}}
\textbf{\bibinfo{volume}{460}},
  \bibinfo{pages}{5-8} (\bibinfo{year}{2006}).

\bibitem{galyam07}
\bibinfo{author}{{Gal-Yam}, A.} \emph{et~al.}
\newblock \bibinfo{title}{{On the Progenitor of SN 2005gl and the nature
    of Type IIn Supernovae}}
\newblock \emph{\bibinfo{journal}{Astrophysical Journal}}
\textbf{\bibinfo{volume}{656}},
  \bibinfo{pages}{372-381} (\bibinfo{year}{2007}).

\bibitem{fransson02}
\bibinfo{author}{{Fransson}, C.} \emph{et~al.}
\newblock \bibinfo{title}{{Optical and Ultraviolet Spectroscopy of
    SN 1995N: Evidence for Strong Circumstellar Interaction}}
\newblock \emph{\bibinfo{journal}{Astrophysical Journal}}
\textbf{\bibinfo{volume}{572}}, 
  \bibinfo{pages}{350--370} (\bibinfo{year}{2002}).

\bibitem{fransson05}
\bibinfo{author}{{Fransson}, C.} \emph{et~al.}
\newblock \bibinfo{title}{{Hubble Space Telescope and Ground-Based 
    Observations of SN 1993J and SN 1998S: CNO Processing in the Progenitors}}
\newblock \emph{\bibinfo{journal}{Astrophysical Journal}}
\textbf{\bibinfo{volume}{622}}, 
  \bibinfo{pages}{991--1007} (\bibinfo{year}{2005}).
 
\bibitem{cooke08}
\bibinfo{author}{{Cooke} J.}
\newblock \bibinfo{title}{{Detecting $z>2$ Type IIn Supernovae}}
\newblock \emph{\bibinfo{journal}{Astrophysical Journal}}
\textbf{\bibinfo{volume}{677}}, 
  \bibinfo{pages}{137--145} (\bibinfo{year}{2008}).

\bibitem{brown08}
\bibinfo{author}{{Brown}, P.} \emph{et~al.}
\newblock \bibinfo{title}{{Ultraviolet Light Curves of Supernovae
    with Swift UVOT}}
\newblock \emph{\bibinfo{journal}{arXiv:0803.1265}}
\textbf{\bibinfo{volume}{}}, 
  \bibinfo{pages}{} (\bibinfo{year}{2008}).

\bibitem{dahlen99}
\bibinfo{author}{{Dahl\'{e}n} T.} { \& Fransson, C.}
\newblock \bibinfo{title}{{Rates and redshift distributions of high-z
    supernovae}}
\newblock \emph{\bibinfo{journal}{Astronomy \& Astrophysics}}
\textbf{\bibinfo{volume}{350}},
  \bibinfo{pages}{345--367} (\bibinfo{year}{1999}).

\bibitem{salpeter55}
\bibinfo{author}{{Salpeter}, E.,~E.}
\newblock \bibinfo{title}{{The Luminosity Function and Stellar Evolution}}
\newblock \emph{\bibinfo{journal}{Astrophysical Journal}}
\textbf{\bibinfo{volume}{121}},
  \bibinfo{pages}{161-167} (\bibinfo{year}{1955}).

\bibitem{vandokkum08}
\bibinfo{author}{{van Dokkum}, P. G.} 
\newblock \bibinfo{title}{{Evidence of Cosmic Evolution of the Stellar 
    Initial Mass Function}}
\newblock \emph{\bibinfo{journal}{Astrophysical Journal}}
\textbf{\bibinfo{volume}{674}},
  \bibinfo{pages}{29--50} (\bibinfo{year}{2008}).

\bibitem{chary08}
\bibinfo{author}{{Chary}, R-R.}
\newblock \bibinfo{title}{{The Stellar Initial Mass Function at the Epoch 
    of Reionization}}
\newblock \emph{\bibinfo{journal}{Astrophysical Journal}}
\textbf{\bibinfo{volume}{680}},
  \bibinfo{pages}{32--40} (\bibinfo{year}{2008}).

\bibitem{dave08}
\bibinfo{author}{{Dav\'{e}}, R.} 
\newblock \bibinfo{title}{{The galaxy stellar mass-star formation rate 
    relation: evidence for an evolving initial mass function}}
\newblock \emph{\bibinfo{journal}{Mon. Not. R. Astron. Soc.}}
\textbf{\bibinfo{volume}{385}},
  \bibinfo{pages}{147--160} (\bibinfo{year}{2008}).

\bibitem{steidel03}
\bibinfo{author}{{Steidel}, C. C.} \emph{et~al.}
\newblock \bibinfo{title}{{Lyman Break Galaxies at Redshift $z\sim3$: 
    Survey Description and Full Data Set}}
\newblock \emph{\bibinfo{journal}{Astrophysical Journal}}
\textbf{\bibinfo{volume}{592}}, 
  \bibinfo{pages}{728--754} (\bibinfo{year}{2003}).

\bibitem{steidel04}
\bibinfo{author}{{Steidel}, C. C.} \emph{et~al.}
\newblock \bibinfo{title}{{A Survey of Star-Forming Galaxies in the 
    $1.4<z<2.5$ Redshift Desert: Overview}}
\newblock \emph{\bibinfo{journal}{Astrophysical Journal}}
\textbf{\bibinfo{volume}{604}}, 
  \bibinfo{pages}{534--550} (\bibinfo{year}{2004}).

\bibitem{cooke05}
\bibinfo{author}{{Cooke}, J.} \emph{et~al.}
\newblock \bibinfo{title}{{Survey for Galaxies Associated With $z\sim3$ 
    Damped Ly-$\alpha$ Systems. I. Spectroscopic Calibration of $u'$BVRI
    Photometric Selection}}
\newblock \emph{\bibinfo{journal}{Astrophysical Journal}}
\textbf{\bibinfo{volume}{621}}, 
  \bibinfo{pages}{596--614} (\bibinfo{year}{2005}).
 
\bibitem{lefevre03}
\bibinfo{author}{{Le F{\`e}vre}, O.} \emph{et~al.}
\newblock \bibinfo{title}{{Virmos-VLT deep survey (VVDS)}}
\newblock \emph{\bibinfo{journal}{SPIE}}
\textbf{\bibinfo{volume}{4834}}, 
  \bibinfo{pages}{173--182} (\bibinfo{year}{2003}).

\bibitem{neill06}
\bibinfo{author}{{Neill}, J. D.} \emph{et~al.}
\newblock \bibinfo{title}{{The Type Ia Supernova Rate at z~0.5 from the 
    Supernova Legacy Survey}}
\newblock \emph{\bibinfo{journal}{Astronomical Journal}}
\textbf{\bibinfo{volume}{132}},
  \bibinfo{pages}{1126--1145} (\bibinfo{year}{2006}).

\bibitem{faber03}
\bibinfo{author}{{Faber}, S. M.} \emph{et~al.}
\newblock \bibinfo{title}{{The DEIMOS spectrograph for the Keck II
    Telescope: integration and testing}}
\newblock \emph{\bibinfo{journal}{SPIE}}
\textbf{\bibinfo{volume}{4841}}, 
  \bibinfo{pages}{1657--1669} (\bibinfo{year}{2003}).

\bibitem{oke95}
\bibinfo{author}{{Oke}, J. B.} \emph{et~al.}
\newblock \bibinfo{title}{{The Keck Low-Resolution Imaging Spectrometer}}
\newblock \emph{\bibinfo{journal}{Pub. of the Astron. Soc. of the Pacific}}
\textbf{\bibinfo{volume}{107}}, 
  \bibinfo{pages}{375--385} (\bibinfo{year}{1995}).

\bibitem{mccarthy98}
\bibinfo{author}{{McCarthy}, J. K.} \emph{et~al.}
\newblock \bibinfo{title}{{Blue channel of the Keck low-resolution 
    imaging spectrometer}}
\newblock \emph{\bibinfo{journal}{SPIE}}
\textbf{\bibinfo{volume}{3355}}, 
  \bibinfo{pages}{81--91} (\bibinfo{year}{1998}).

\bibitem{riess04}
\bibinfo{author}{{Riess}, A. G.} \emph{et~al.}
\newblock \bibinfo{title}{{Identification of Type Ia Supernovae at Redshift
    1.3 and Beyond With the Advanced Camera for Surveys on the Hubble Space
    Telescope}}
\newblock \emph{\bibinfo{journal}{Astrophysical Journal Letters}}
\textbf{\bibinfo{volume}{600}}, 
  \bibinfo{pages}{163--166} (\bibinfo{year}{2004}).

\bibitem{shapley03}
\bibinfo{author}{{Shapley}, A. E.} \emph{et~al.}
\newblock \bibinfo{title}{{Rest-frame Ultraviolet Spectra of 
    $z\sim3$ Lyman Break Galaxies}}
\newblock \emph{\bibinfo{journal}{Astrophysical Journal}}
\textbf{\bibinfo{volume}{558}},
  \bibinfo{pages}{65--89} (\bibinfo{year}{2003}).

\end{thebibliography}

\clearpage

\begin{thebibliography}{10}
\expandafter\ifx\csname url\endcsname\relax
  \def\url#1{\texttt{#1}}\fi
\expandafter\ifx\csname urlprefix\endcsname\relax\def\urlprefix{URL }\fi
\providecommand{\bibinfo}[2]{#2}
\providecommand{\eprint}[2][]{\url{#2}}

\bibitem[31]{landolt92} Landolt, A. ~UBVRI photometric standard in the
  magnitude range 11.5-16.0 around the celestial equator
  \emph{Astronomical Journal} \textbf{104}, {340--491} (1992).

\bibitem[32]{bertin96} Bertin, E. \& Arnouts, S. ~SExtractor: Software
  for source extraction \emph{Astron. \& Astroph. Supp. Series}
  \textbf{117}, 393--404 (1996).

\bibitem[33]{madau95} Madau, P. ~Radiative transfer in a clumpy
  universe: The colors of high-redshift galaxies \emph{Astrophysical
  Journal} \textbf{441}, 18--27 (1995).

\bibitem[34]{immler06} Immler, S. \& Pooley, D. ~Swift Observations of
  SN 2006bv in UGC 7848 \emph{The Astronomer's Telegram} \textbf{802},
  1--1 (2006).

\bibitem[35]{immler07} Immler, S. \& Brown, P.~J.  ~Swift Observations
  of SN 2007bb \emph{The Astronomer's Telegram} \textbf{1053}, 1--1
  (2007).

\bibitem[36]{adelberger04} Adelberger, K. L. \emph{et~al.} ~Optical
  Selection of Star-forming Galaxies at Redshifts $1 < z < 3$
  \emph{Astrophysical Journal} \textbf{607}, 226--240 (2004).

\bibitem[37]{adelberger03} Adelberger, K.~L. \emph{et~al.} ~Galaxies
  and Intergalactic Matter at Redshift $z\sim3$: Overview
  \emph{Astrophysical Journal} \textbf{584}, 45--75 (2003).

\bibitem[38]{filippenko89} Filippenko, A.~V. ~The 'Seyfert 1' optical
  spectra of the type II supernovae 1987F and 1988I \emph{Astronomical
  Journal} \textbf{97}, 726--734 (1989).

\bibitem[39]{leonard00} Leonard, D.~L. \emph{et~al.} ~Evidence for
  Asphericity in the Type IIN Supernova SN 1998S \emph{Astrophysical
  Journal} \textbf{536}, 239--254 (2000).

\bibitem[40]{vandenberk01} Vanden Berk, D.~E. \emph{et~al.} ~Composite
  Quasar Spectra from the Sloan Digital Sky Survey \emph{Astronomical
  Journal} \textbf{122}, 549--564 (2001).

\bibitem[41]{glikman07} Glikman, E. \emph{et~al.} ~Discovery of Two
  Spectroscopically Peculiar, Low-Luminosity Quasars at $z\sim4$
  \emph{Astrophysical Journal Letters} \textbf{663}, 73--76 (2007).

\end{thebibliography}
\end{document}